\documentclass[useAMS,usegraphicx,fleqn,usenatbib,letterpaper]{mn2e} 
\usepackage{graphicx,epsfig}
\usepackage{color}
\usepackage{url}
\usepackage{ulem}
\usepackage{amssymb}
\usepackage{amsmath}
\usepackage{wasysym}
\usepackage{subfigure}
\usepackage{multirow}
\usepackage{natbib}
\usepackage{pifont}
 \usepackage{times}

\def\apj{{\em ApJ}}
\def\apjs{{\em ApJS}}
\def\apjl{{\em ApJ}}
\def\aap{{\em A\&A}}
\def\aj{{\em AJ}}

\def\mnras{{\em MNRAS}}

\def\araa{Annual Review of Astron and Astrophys}
 
\def\pasp{{\em PASP}}
\def\pasa{{\em Publ. Astron. Soc. Australia}}
	
\def\nar{{\em New Astronomy Reviews}}	
\def\procspie{{\em Proc. SPIE}}	
\newcommand{\lsun}{\mbox{\ensuremath{\,\rmn{L_{\odot}}}}}       
\newcommand{\msun}{\mbox{\ensuremath{\,\rmn{M_{\odot}}}}}       
\newcommand{\lunits}{\ensuremath{\,\rmn{erg\,s^{-1}\,Hz^{-1}}}} 
\newcommand{\rlunits}{\ensuremath{\,\rmn{erg\,s^{-1}}}}         
\newcommand{\funits}{\ensuremath{\,\rmn{erg\,s^{-1}\,cm^{-2}}}} 
\newcommand{\phunits}{\ensuremath{\,\rmn{photons\,s^{-1}\,cm^{-2}}}}

\newcommand{\h}{\ensuremath{\rmn{^h}}}                           
\newcommand{\m}{\ensuremath{\rmn{^m}}}                           
                           
\newcommand{\kms}{\,\mbox{\ensuremath{\rmn{km\,s^{-1}}}}}        
\newcommand{\mum}{\,\mbox{\ensuremath{\rmn{\mu m}}}}             
\newcommand{\mjy}{\,\mbox{\ensuremath{\rmn{mJy}}}}               
           
\newcommand{\jyb}{\,\mbox{\ensuremath{\rmn{Jy/beam}}}}           
\newcommand{\mjyb}{\,\mbox{\ensuremath{\rmn{mJy/beam}}}}         
\newcommand{\mujyb}{\,\mbox{\ensuremath{\rmn{\mu Jy/beam}}}}     
     
\newcommand{\hii}{\mbox{H{\footnotesize{II}}}}                   
\newcommand{\hi}{\mbox{H{\footnotesize{I}}}}

\newcommand{\lir}{\ensuremath{L_{\rmn{IR}}}}                     
                   
\newcommand{\ledd}{\ensuremath{L_{\rmn{Edd}}}}                   
\newcommand{\lemi}{\ensuremath{L_{\rmn{emitted}}}}               
\newcommand{\qjet}{\ensuremath{Q_{\rmn{jet}}}}                   
\newcommand{\lbolagn}{\ensuremath{L_{\rmn{\mbox{\tiny Bol}}}^{\rmn{\mbox{\tiny AGN}}}}}
\newcommand{\liragn}{\ensuremath{L_{\rmn{\mbox{\tiny IR}}}^{\rmn{\mbox{\tiny AGN}}}}}  
\newcommand{\mbh}{\ensuremath{M_{\rmn{BH}}}}                                        
\newcommand{\alphcx}{\ensuremath{\alpha_{\rmn{\mbox{\tiny C-X}}}}} 
\newcommand{\alphlc}{\ensuremath{\alpha_{\rmn{\mbox{\tiny L-C}}}}} 
\newcommand{\tb}{\ensuremath{T_{\rmn{B}}}}                         
\defcitealias{rocc12a}{Paper~I}
\title[The AGN activity in IC\,883]
{Unveiling the AGN in IC\,883: discovery of a parsec-scale radio jet}

\author[C.~Romero-Ca\~nizales, et al.]{
C.\ Romero-Ca\~nizales,$^{1, 2, 3}$\thanks{E-mail: cristina.romero.fdi@mail.udp.cl}
A.\ Alberdi,$^{4}$
C. Ricci,$^{2}$
P.\ Ar\'evalo,$^{5}$
M.\ \'A.\ P\'erez-Torres,$^{4, 6}$\thanks{Visiting Scientist}
\newauthor 
J.\ E. Conway,$^{7}$
R.\ J.\ Beswick,$^{8}$
M.\ Bondi,$^{9}$
T.\ W.\ B.\ Muxlow,$^{8}$
M.\ K.\ Argo,$^{8,10}$
\newauthor 
F.\ E.\ Bauer,$^{2, 1, 11, 12}$
A.\ Efstathiou,$^{13}$
R.\ Herrero-Illana,$^{4,14}$
S.\ Mattila$^{15}$ and
S.\ D.\ Ryder$^{16}$\\
$^{1}$Millennium Institute of Astrophysics, Chile \\
$^{2}$Instituto de Astrof\'{\i}sica, Facultad de F\'{\i}sica, Pontificia Universidad Cat\'olica de Chile, Casilla 306, Santiago 22, Chile \\
$^{3}$N\'ucleo de Astronom\'{\i}a de la Facultad de Ingenier\'{\i}a y Ciencias, Universidad Diego Portales, Av. Ej\'ercito 441, Santiago, Chile  \\
$^{4}$Instituto de Astrof\'{\i}sica de Andaluc\'{\i}a -- CSIC, PO Box 3004, 18080 Granada, Spain \\
$^{5}$Instituto de F\'{\i}sica y Astronom\'{\i}a, Facultad de Ciencias, Universidad de Valpara\'{\i}so, Gran Breta\~na No. 1111, Playa Ancha, Valpara\'{\i}so, Chile \\
$^{6}$Departamento de F\'isica Teorica, Facultad de Ciencias, Universidad de Zaragoza, Spain \\
$^{7}$Department of Earth and Space Sciences, Chalmers University of Technology, Onsala Space Observatory, 439 92, Onsala, Sweden \\
$^{8}$Jodrell Bank Centre for Astrophysics, The University of Manchester, Oxford Rd, Manchester M13 9PL, UK \\
$^{9}$Osservatorio di Radioastronomia-INAF, Bologna, via P. Gobetti 101, 40129 Bologna, Italy \\
$^{10}$Jeremiah Horrocks Institute, University of Central Lancashire, Preston, PR1 2HE, UK \\
$^{11}$Centro de Astroingenier\'{\i}a, Facultad de F\'{i}sica, Pontificia Universidad Cat\'olica de Chile, Casilla 306, Santiago 22, Chile \\
$^{12}$Space Science Institute, 4750 Walnut Street, Suite 205, Boulder, CO 80301, USA \\
$^{13}$School of Sciences, European University Cyprus, Diogenes Street, Engomi, 1516 Nicosia, Cyprus \\
$^{14}$European Southern Observatory (ESO), Alonso de C\'ordova 3107, Vitacura, Casilla 19001, Santiago de Chile, Chile \\
$^{15}$Tuorla Observatory, Department of Physics and Astronomy, University of Turku, V\"ais\"al\"antie 20, FI- 21500 Piikki\"o, Finland \\
$^{16}$Australian Astronomical Observatory, PO Box 915, North Ryde, NSW 1670, Australia}

\begin{document}

\date{Accepted 2017 January 23. Received 2017 January 23; in original form 2016 June 29}
\pagerange{\pageref{firstpage}--\pageref{lastpage}} \pubyear{2017}

\maketitle

\label{firstpage} 

\begin{abstract}
IC\,883 is a luminous infrared galaxy (LIRG) classified as a starburst-active galactic nucleus
(AGN) composite. In a previous study we detected a low-luminosity AGN (LLAGN) radio candidate. 
Here we report on our radio follow-up at three frequencies which provides direct and unequivocal
evidence of the AGN activity in IC\,883. Our analysis of archival X-ray data, together with the 
detection of a transient radio source with luminosity typical of bright supernovae, give further 
evidence of the ongoing star formation activity, which dominates the energetics of the system. At 
sub-parsec scales, the radio nucleus has a core-jet morphology with the jet being a newly ejected 
component showing a subluminal proper motion of 0.6--1\,$c$. The AGN contributes less than two per 
cent of the total IR luminosity of the system. The corresponding Eddington factor is $\sim10^{-3}$, 
suggesting this is a low-accretion rate engine, as often found in LLAGNs. However, its high 
bolometric luminosity ($\sim10^{44}$\rlunits{}) agrees better with a normal AGN. This apparent 
discrepancy may just be an indication of the transition nature of the nucleus from a system 
dominated by star-formation, to an AGN-dominated system. The nucleus has a strongly inverted 
spectrum and a turnover at $\sim4.4$\,GHz, thus qualifying as a candidate for the least luminous 
($L_{\rmn{5.0\,GHz}}\sim 6.3\times10^{28}$\,\lunits{}) and one of the youngest 
($\sim3\times10^3$\,yr) gigahertz-peaked spectrum (GPS) sources.  If the GPS origin for the 
IC\,883 nucleus is confirmed, then advanced mergers in the LIRG category are potentially key 
environments to unveil the evolution of GPS sources into more powerful radio galaxies.
\end{abstract}


%%%%%%%%%%%%%%%%%%%%%%%%%%%%%%%%%%%%%%%%%%%%%%%%%%

%%%%%%%%%%%%%%%%% BODY OF PAPER %%%%%%%%%%%%%%%%%%

\section{Introduction}\label{sec:intro}

One of the mechanisms in the Universe able to trigger bursts of star formation, as well as the onset of 
an active galactic nucleus (AGN), is the merger of gas-rich galaxies \citep[e.g.,][]{hopkins06,dimatteo07}.
These systems are a seed for galaxy evolution and are thought to produce luminous and ultraluminous infrared 
(IR: 8--1000\,\mum) galaxies \citep[LIRGs: $\lir>10^{11}$\,\lsun{}; ULIRGs: $\lir>10^{12}$\,\lsun{};][]{sanders96}. 
Owing to (U)LIRGs nature, both the AGN and the active star formation therein are deeply embedded in dust, and hence 
obscured. Radio observations provide us with an extinction-free view of the merger products, and thus allow
us to study their interaction, their evolution and their influence on their hosts.

IC\,883, also known as UGC\,8387, I\,Zw\,056 and Arp\,193, is one of the closest ($D=100$\,Mpc) LIRGs, 
with an IR luminosity of $\sim4.7\times10^{11}$\,\lsun{} \citep[]{sanders03}. The system 
consists of a $\sim4$\,kpc rotating ring of molecular gas and dust observed edge-on, and two large 
tidal tails (one protruding orthogonally from the centre and another extending along the main 
body towards the South), which indicate the occurrence of previous merger episodes 
\citep[e.g.,][]{smith95,downes98,scoville00}. \citet{clemens04} found kinematic differences in 
\hi{} absorption and $^{12}$CO emission which they interpret as evidence of an outflow of atomic 
gas perpendicular to the ring, though they also note inflow (or other non-circular motion) could 
account for the velocity differences. It is possible that returning tidal material at somewhat 
larger distances, but seen in projection against the nucleus, could also be the origin of the 
kinematic disparity between the $^{12}$CO and \hi{}.

\citet{veilleux95} classified this galaxy as a low-ionisation nuclear emission-line region 
(LINER) based on optical spectroscopy. Posterior analysis suggested its classification as 
a starburst--AGN composite \citep{yuan10}. The thermal nature 
of the soft X-ray emission, the IR colours indicating thermal emission from cold dust
\citep[][and references therein]{modica12}, the steep global radio spectral index ($-0.64$)
and the far-IR/radio flux ratio \citep[$q=2.28$;][]{condon91}, among other characteristics, 
provide solid evidence of active star formation in the system. In fact, IC\,883 lies close 
to the well known starburst galaxy M82 in the diagnostic plot of the 6.2\,\mum{} Polycyclic 
Aromatic Hydrocarbon  vs the 9.7\,\mum{} silicate strength \citep[with values of 0.6 and $-1$, 
respectively;][]{stierwalt13} proposed by \citet{spoon07}. Support for the presence 
of an AGN in IC\,883 comes from the [Ne {\small V}] 14.32\,\mum{} emission line detection 
in the heavily obscured nuclear region \citep{dudik09}, and the detection of compact radio 
sources at milli-arcsec resolution \citep[][]{lonsdale93,smith98,parra10}. The AGN has however 
remained elusive in X-rays and mid-IR continuum observations, where AGNs are typically 
uncovered, and hence its presence has been judged uncertain \citep[e.g.,][]{asmus15}. We note 
that the starburst-AGN classification means solely that there is both emission from a starburst 
and merger-driven shocks, which together can mimic strong AGN emission \citep{rich14}. 

\citet{papadopoulos14} studied the spectral line energy distribution (SLED) of CO lines in IC\,883, 
plus emission lines of dense gas tracers such as high-{\it J} HCN transitions, and found that the system 
basically lacks a dense molecular gas component ($n>10^4$\,cm$^{-3}$). This is a somewhat unexpected
result for a system which is presumably undergoing a starburst.

In \cite{rocc12a}, henceforth \citetalias{rocc12a}, we presented radio observations of the nuclear 
and circumnuclear regions of IC\,883 using very long baseline interferometry (VLBI) facilities and 
the electronic Multi-Element Remotely Linked Interferometer Network (e-MERLIN) array. Our observations 
revealed the presence of at least six non-thermal compact components (labelled as components A1 to A6) 
within the innermost 100\,pc diameter nuclear region of IC\,883 (named component A), likely constituting 
a supernova (SN) factory in coexistence with a low-luminosity AGN (LLAGN) candidate (component A1). We 
found that the AGN candidate is powering the radio emission at both circumnuclear and nuclear scales, 
as seen with the e-MERLIN and European VLBI Network (EVN) arrays, unlike the scenario depicted at other 
wavelengths, where the starburst represents the major contribution to the global emission.

In \citetalias{rocc12a} we also modelled the IR spectral energy distribution (SED) and found that the 
contribution from a putative AGN to the IR emission is no more than 10 per cent that of the starburst, 
albeit with inherent large uncertainties predicted by our almost edge-on view of the torus.

In this paper we present recent VLBI radio observations made with the EVN, as well as archival X-ray 
data, aiming at confirming the presence of an AGN in IC\,883, and at understanding its nature. We also 
present archival Karl G.{} Jansky Very Large Array (VLA) data which has a resolution comparable to that 
of our e-MERLIN observations reported in \citetalias{rocc12a}. Our study is made within the framework 
of the e-MERLIN legacy project Luminous InfraRed Galaxy Inventory (LIRGI; PIs: J. Conway \& M. \'A. 
P\'erez-Torres)\footnote{http://lirgi.iaa.es/}, which will offer a radio complement to the Great 
Observatories All-sky LIRG Survey \citep[GOALS;][]{armus09} NASA program. LIRGI pursues the ambitious 
goal of characterising the nuclear and circumnuclear radio emission of a statistically significant 
sample of 42 of the most luminous northern LIRGs.

We organise the manuscript as follows. In \S \ref{sec:obs} we describe the radio and X-ray observations 
used, as well as the data reduction process. In \S \ref{sec:results} we provide maps and measurements 
from our data. In \S \ref{sec:discussion} we investigate the nature of the IC\,883 nucleus based on its 
morphology, flux density variability, its radio SED and the X-ray emission. We recap in \S \ref{sec:fin} 
with our final conclusion.

\section{Observations: description and data reduction}\label{sec:obs}

\subsection{VLBI radio follow-up}

With the aim of characterising the nuclear emission in IC\,883 we started an observing program using the EVN 
under project code ER030 (PI: C. Romero-Ca\~nizales). The program consisted of three yearly epochs from 2012 
to 2014 (see details in Table \ref{tab:er030}) at the frequency bands centred at $\nu = 8.4$\,GHz ($\lambda = 
3.6$\,cm, or {\it X}-band), $\nu = 4.9$\,GHz ($\lambda = 6.1$\,cm, or {\it C}-band) and $\nu =1.6$\,GHz 
($\lambda = 18.7$\,cm, or {\it L}-band), using the following stations (location, diameter): Ef-Effelsberg 
(DE, 100\,m), Wb-Westerbork array (NL, 14$\times$25\,m), Jb1-Lovell (UK, 76\,m), Jb2-MK\,II (UK, 25\,m), 
On-Onsala (SE, 20\,m at $\lambda < 5$\,cm, and 25\,m at $\lambda \ga 5$\,cm), Mc-Medicina (IT, 32\,m), Nt-Noto 
(IT, 32\,m), Tr-Torun (PL, 32\,m), Ys-Yebes (ES, 40\,m), Sv-Svetloe (RU, 32\,m), Zc-Zelenchukskaya 
(RU, 32\,m), Bd-Badary (RU, 32\,m), Ur-Nanshan (CN, 25\,m), Sh-Sheshan (CN, 25\,m), Hh-Hartebeesthoek 
(SA, 26\,m) and Ro-Robledo (ES, 70\,m). 

Each observing segment (see Table \ref{tab:er030}) lasted 4\,hr in total, from which 2.5\,hr were spent on 
target. An integration time of 2\,sec was used in all the epochs, except for segments D and G, where we used 
1\,sec to obtain a similar field of view as for the other frequencies in the same epochs. We used $8\times16$\,MHz 
sub-bands, each with dual polarisation. In the 2012 epoch the observations at 4.9 and 1.6\,GHz were carried out under 
project EP076 (segments C and D), and subject to a correlation flaw which made those observations unusable. 

We reduced the data in the NRAO Astronomical Image Processing System ({\sc aips}). Taking as a 
starting point the EVN pipeline products, we improved the calibration considering ionospheric 
corrections, radio interference removal, and phase and amplitude self-calibration on the phase 
reference source, J1317$+$3425. The average peak intensities of this calibrator in the different 
bands were $0.24\pm0.01$\jyb{}, $0.32\pm0.02$\jyb{} and $0.19\pm0.01$\jyb{}, with a variability 
in the peak intensity among epochs of up to 11, 18 and 9 per cent at 8.4, 4.9 and 1.6\,GHz, 
respectively. In all segments, we used J1159$+$2914 as a fringe finder.

%%%%%%%%%%%%%%%%%%%%%%%%%%%%%%%%%%%%%%%%%%%%%%%%%%%%%%%%%%%%%%%%%%%%%%%%%%%%%%%%%%%%%%%%%
\begin{table*}
\centering
\begin{minipage}{168mm}
\caption{\protect{Observation setup for EVN project ER030. Stations in italics either did not observe,
or were completely lost due to weather/technical problems.}} \label{tab:er030}
\begin{tabular}{cccl} \hline
ER030    & Central & Observing & \multicolumn{1}{c}{Participating}  \\
segment  & frequency (GHz)     & date      & \multicolumn{1}{c}{stations}       \\
    \hline
- & 8.4 & 02-Nov-2012 & Ef, Wb, On, Mc, Nt, Ys, Sv, Zc, Bd, Ur, Sh, Hh, Ro \\
B & 4.9 & 24-Oct-2013 & Ef, Wb, Jb2, {\it On}, Mc, Nt, {\it Tr}, Ys, Sv, Zc, Bd, {\it Ur}, {\it Sh}, Hh   \\
C & 1.6 & 31-Oct-2013 & {\it Ef}, Wb, {\it Jb1}, On, Mc, {\it Nt}, Tr, Sv, Zc, Bd, {\it Ur}, {\it Sh}, Hh \\
D & 8.4 & 04-Nov-2013 & {\it Ef}, Wb, {\it On}, Mc, {\it Nt}, Ys, Sv, Zc, Bd, Ur, {\it Sh}, Hh \\
E & 4.9 & 22-Oct-2014 & Ef, Wb, Jb1, On, Nt, Tr, Ys, Sv, Zc, Bd, Sh, Hh    \\
F & 1.6 & 30-Oct-2014 & Ef, Wb, Jb1, On, {\it Nt}, Tr, Sv, Zc, Bd, Sh, Hh  \\
G & 8.4 & 04-Nov-2014 & Ef, Wb, On, {\it Nt}, Ys, Sv, Zc, Bd, Sh, Hh       \\
    \hline
   \end{tabular} 
 \end{minipage}
\end{table*}
%%%%%%%%%%%%%%%%%%%%%%%%%%%%%%%%%%%%%%%%%%%%%%%%%%%%%%%%%%%%%%%%%%%%%%%%%%%%%%%%%%%%%%%%%

Imaging using different weighting regimes was performed both in {\sc aips} and the Caltech 
imaging programme {\sc difmap} \citep{shepherd95}, to assess the robustness of our results.

Throughout this paper, the uncertainties of the peak intensities and flux densities include a 
5 per cent calibration uncertainty added in quadrature to the r.m.s. measured in the image. 
For the flux densities we also consider a multiplying factor for the r.m.s. term, equal to the 
number of beams covering the emitting region.

\subsection{VLA archival observations}\label{sec_vla}

We reduced archival VLA data at 32.5\,GHz, which has a resolution comparable to our previously 
reported e-MERLIN observations at 7.9\,GHz \citepalias{rocc12a}. Comparing the radio emission 
at these two frequencies can offer further understanding on the nature of the nucleus, given 
that the LLAGN candidate in IC\,883 dominates the radio emission at both circumnuclear and 
nuclear scales.

IC\,883 was observed at 32.5\,GHz (central frequency) with the VLA on 12, 14 and 15 June 2011 
under project AL746 (PI: A. K. Leroy), when the array was in its most extended configuration.
These observations have been reported by \citet{barcos16}.
 
We used the Common Astronomy Software Application \citep[{\sc casa};][]{mcmullin07} to reduce 
the data. The three datasets were processed independently using the VLA pipeline and removing 
obvious radio frequency interference features. The phase reference source was J1317$+$3425, as 
in our EVN observations. The peak intensity did not vary among datasets and we thus proceeded 
to combine them into a single dataset that allowed us to make a deeper image. 

\subsection{X-ray archival data}

\subsubsection{{\it XMM-Newton} observations}

IC\,883 was observed by {\it XMM-Newton} \citep{jansen01} on 09 January 2001 for 21\,ks 
(ObsID 0093640401, PI: F. Bauer). The observations were published by \citet{carrera07}.
The PN camera \citep{struder01} was operated in Extended Full Frame mode and the MOS1 and 
MOS2 cameras \citep{turner01} in Prime Full Window, all using the Medium filter. We processed 
the data using the latest version of {\sc sas} available (14.0.0). Source and background events 
were extracted for all detectors using an aperture radius of 40\arcsec{} and selecting 
background regions on a source-free area on the same chip as the target. For the PN 
camera we selected single and double events and for both MOS detectors we selected single, 
double, triple and quadruple events. In all cases only events with quality flag$=0$ were retained. 
The resulting net (source minus background) count rates and exposures were 0.015\,counts/s 
and 19.5\,ks for MOS1, 0.013\,counts/s  and 19.6\,ks for MOS2, and 0.043\,counts/s  and 
13.8\,ks for the PN. We generated response files for each detector using the {\sc rmfgen} and
{\sc arfgen} tasks and grouped the spectra with a minimum of 20 counts per bin using the task 
{\sc specgroup}.

\subsubsection{{\it Chandra} observations}

We retrieved and re-analysed {\it Chandra} \citep{weisskopf00} observations of IC\,883 
(ObsID 7811, PI: D. Sanders) made on UT 20 February 2007, and originally published by 
\citet{iwasawa11}. The observation was carried out with ACIS-S \citep{garmire03} and
lasted for 14.2\,ks. We performed the data reduction using {\sc ciao} v.4.6 and
following the standard procedure. We reprocessed the data with {\sc chandra\_repro} 
and then extracted the spectra using the {\sc specextract} tool.         

To be consistent with the spectral extraction of {\it XMM-Newton}/EPIC PN and MOS, 
which have much lower spatial resolution than {\it Chandra}, the ACIS-S source spectrum 
used for the spectral fitting was extracted from a circular region of 10\,arcsec radius. 
The background spectrum was extracted from a circular region of the same size on the
same CCD, where no other source was detected.

\section{Analysis and Results}\label{sec:results}

\subsection{The pc-scale jet in IC\,883}\label{sec:jet}

\begin{figure}
\centering
 \includegraphics[angle=0, scale=0.45]{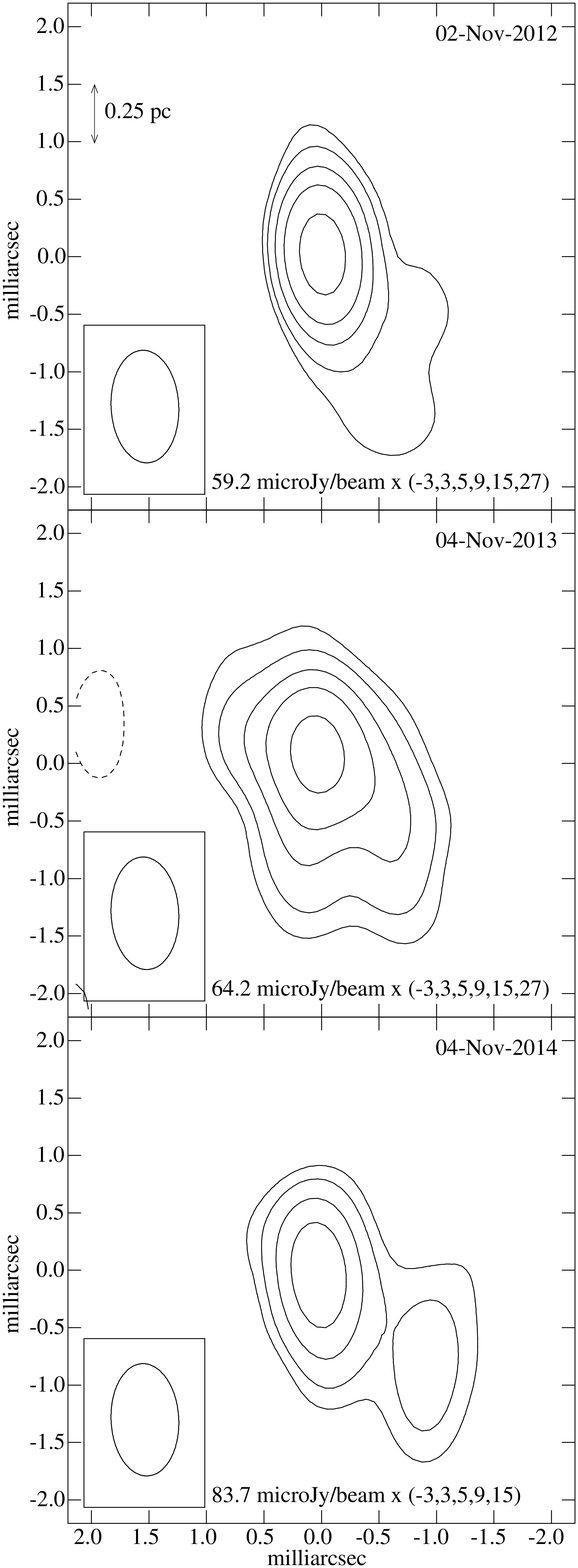} 
 \caption{Contour maps of IC\,883 nuclear region from 8.4\,GHz EVN observations showing the evolution 
  of the radio morphology at a resolution of $0.98\times0.59$\,mas$^2$ at PA$=$2.95\degr{}. The maps 
  are centred at the position of the core: $\alpha_{\rmn{core}}(\rmn{J}2000) =   13\h20\m35\fs3184$, 
  $\delta_{\rmn{core}}(\rmn{J}2000) = 34\degr08\arcmin22\farcs352$. Negative contours appear in dashed 
  style.}\label{fig:core_jet}
\end{figure}

The 8.4\,GHz EVN images benefit from an exquisite angular resolution and reveal for the first 
time a jet-like component emerging from the IC\,883 core (see Fig. \ref{fig:core_jet}). 
In Table \ref{tab:xuni} we show the parameters obtained from the 8.4\,GHz images convolved 
with a beam of $0.98\times0.59$\,mas$^2$ at PA$=$2.95\degr{}. For the flux measurements of the 
core, we considered only the emission within the 9$\sigma$ contour to exclude most of the extended 
emission that is likely associated with the jet (see Fig. \ref{fig:core_jet}). 

%%%%%%%%%%%%%%%%%%%%%%%%%%%%%%%%%%%%%%%%%%%%%%%%%%%%%%%%%%%%%%%%%%%%%%%%%%%%%%%%%%%%%%%%%
\begin{table*}
\centering
{\footnotesize
\begin{minipage}{168mm}
\caption{\protect{Parameters estimated from the 8.4\,GHz EVN images at $0.98\times0.59$\,mas$^2$, 
PA$=$2.95\degr{}.}} \label{tab:xuni}
\begin{tabular}{cccccccc} \hline
Epoch & Component & r.m.s.   &   Size    & Peak intensity & Flux density  & Luminosity   & Brightness temperature   \\
~     &   ~       & (\mujyb) & (mas$^2$) & $P_{8.4}$ (\mjyb)  & $S_{8.4}$ (\mjy) & $L_{8.4}$ ($\times 10^{27}$\lunits{})  & \tb{} ($\times 10^7$K) \\
    \hline
2012 & core       & 59.2 & $< 0.98\times$0.59  & 2.19$\pm$0.12 & 2.19$\pm$0.12 & 26.19$\pm$1.49 & $>$6.64 \\
2013 & ~          & 64.2 & 0.87$\times$0.57    & 2.30$\pm$0.13 & 3.11$\pm$0.25 & 37.25$\pm$2.96 & 11.04$\pm$0.88 \\
2014 & ~          & 83.7 & 0.80$\times$0.31    & 1.86$\pm$0.13 & 2.14$\pm$0.24 & 25.66$\pm$2.81 & 15.15$\pm$1.66 \\
\hline
2014 & jet        & 83.7 & 1.37$\times$0.57    & 0.59$\pm$0.09 & 0.80$\pm$0.09 &  9.54$\pm$1.11 & 1.77$\pm$0.21 \\
    \hline
   \end{tabular} 
 \end{minipage}}
\end{table*}
%%%%%%%%%%%%%%%%%%%%%%%%%%%%%%%%%%%%%%%%%%%%%%%%%%%%%%%%%%%%%%%%%%%%%%%%%%%%%%%%%%%%%%%%%
%%%%%%%%%%%%%%%%%%%%%%%%%%%%%%%%%%%%%%%%%%%%%%%%%%%%%%%%%%%%%%%%%%%%%%%%%%%%%%%%%%%%%%%%%%%%%%%%%%%%%%%%%%%%%%%%%% 
%%%%%%%%%%%%%%%%%%%%%%%%%%%%%%%%%%%%%%%%%%%%%%%%%%%%%%%%%%%%%%%%%%%%%%%%%%%%%%%%%%%%%%%%%%%%%%%%%%%%%%%%%%%%%%%%%% 

In the 2013 and 2014 epochs the core appears slightly extended toward the direction opposite to 
the movement of the jet, with the largest morphology distortion in the 2013 epoch. This could hint 
at the existence of a counter-jet, which we cannot directly detect, probably due to the effects of 
Doppler de-boosting. However, we acknowledge that the 2013 observations were subject to technical 
problems which affected directly the {\it uv}-coverage of the source, posing difficulties for the 
image reconstruction and lowering its fidelity. 

Only in the 2014 epoch is it possible to disentangle the jet component from the core, and we detect 
these two distinct components with peak positions 
$\alpha_{\rmn{core}}(\rmn{J}2000) = 13\h20\m35\fs3184$, 
$\delta_{\rmn{core}}(\rmn{J}2000) = 34\degr08\arcmin22\farcs352$ 
and $\alpha_{\rmn{jet}}(\rmn{J}2000) = 13\h20\m35\fs3183$, 
$\delta_{\rmn{jet}}(\rmn{J}2000) = 34\degr08\arcmin22\farcs351$. 
The jet component is thus at a projected distance of 0.62\,pc South-West from the core, and 
at a position angle (PA) of $\sim215\degr{}$, which is almost perpendicular to the IC\,883 
ring structure seen at large scales with e-MERLIN at PA$=$130\degr{}, measured from East to 
North \citepalias[][and references therein]{rocc12a}. 

IC\,883 was also observed at 8.4\,GHz on 15 May 2011 with the very long baseline array (VLBA), 
as reported in \citetalias{rocc12a}. The angular resolution of this observation was twice that 
of the later EVN epochs, hence both the core and any possible jet component would have been 
contained within the same beam. Hence, the use of the VLBA observations as a zero-epoch is not
straightforward. However, we note that in 2011, the flux density was 4.38$\pm$0.28\,mJy 
\citepalias{rocc12a}, and by 2012 it had halved. We infer that the 2014 jet component must have 
been ejected some time between 15 May 2011 and 2 November 2012, thus providing a lower limit to 
its apparent speed of 0.6\,$c$.

Using the 2012 epoch as a reference, we find that the jet has moved at approximately 1.0\,$c$. 
This is an upper limit to its apparent speed, as in the 2012 epoch we already see some extended 
emission from the core towards the direction where the jet component is detected in 2014. 

\subsection{Radio evolution of the IC\,883 nuclear region}\label{sec:radevol}

IC\,883 shows compact structure at both 4.9 and 1.6\,GHz as observed with the EVN. We have 
created images for all observations using a common convolving beam of $7\times6$\,mas$^2$, 
which is the largest synthesised beam obtained in any epoch (corresponding to the 1.6\,GHz 
observations in 2014 imaged with uniform weighting). This allows us to obtain flux densities 
from similar regions and eases the interpretation of the spectral index behaviour, although 
at the expense of increasing the r.m.s. Typically the use of natural weighting improves the 
r.m.s. whilst it provides slightly lower resolution. Here, in addition to the use of natural
weighting, we have also used a much larger convolving beam than that achieved with only 
natural weighting (especially at 8.4\,GHz). This is effectively tapering the data and 
consequently, it results in a larger r.m.s. We use $\alpha$ to denote spectral indices (defined 
by $S_{\nu}\propto \nu^{\alpha}$). The results are shown in Table \ref{tab:lcx_matched}.

%%%%%%%%%%%%%%%%%%%%%%%%%%%%%%%%%%%%%%%%%%%%%%%%%%%%%%%%%%%%%%%%%%%%%%%%%%%%%%%%%%%%%%%%%
\begin{table*}
\centering
{\footnotesize
\begin{minipage}{168mm}

\caption{\protect{Estimated parameters from the matched-beam EVN images of the nucleus 
                  at $7\times6$\,mas$^2$.}} \label{tab:lcx_matched}
\begin{tabular}{cccccccc} \hline
Epoch-Frequency   & r.m.s.   & Deconvolved     & Peak intensity & Flux density  & Luminosity - $L_{\nu}$  & Brightness temp.  & Spectral  \\
year - $\nu$ (GHz) & (\mujyb) & sizes (mas$^2$) &  $P_{\nu}$ (\mjyb) & $S_{\nu}$ (\mjy) &  ($\times 10^{27}$\lunits{}) & \tb{} ($\times 10^7$K) & indices \\
    \hline
2012 -- 8.4 & 77.0 & $< 7\times$6     & 2.66$\pm$0.15 & 2.66$\pm$0.15 & 31.86$\pm$1.84 & $>$0.11 \\
\hline
2013 -- 8.4 & 75.5 & 1.68$\times$0.80 & 4.65$\pm$0.24 & 4.78$\pm$0.31 & 57.19$\pm$3.76 & 6.22$\pm$0.41 & \\[-1ex]
2013 -- 4.9 & 42.2 & $< 7\times$6     & 4.30$\pm$0.22 & 4.30$\pm$0.22 & 51.46$\pm$2.62 & $>$0.51       & \raisebox{1.5ex}{\alphcx=0.20 $\pm$ 0.16}\\[-1ex]
2013 -- 1.6 & 26.0 & 3.97$\times$2.63 & 0.33$\pm$0.03 & 0.41$\pm$0.03 & 4.89$\pm$0.40  & 1.86$\pm$0.15 & \raisebox{1.5ex}{\alphlc=2.09 $\pm$ 0.09}\\
\hline
2014 -- 8.4 & 96.8 & 4.18$\times$1.14 & 4.79$\pm$0.26 & 5.71$\pm$0.39 & 68.29$\pm$4.71 & 2.09$\pm$0.14 & \\[-1ex]
2014 -- 4.9 & 70.0 & 2.90$\times$1.89 & 4.62$\pm$0.24 & 5.26$\pm$0.34 & 62.99$\pm$4.03 & 4.82$\pm$0.31 & \raisebox{1.5ex}{\alphcx=0.15 $\pm$ 0.18}\\[-1ex]
2014 -- 1.6 & 15.0 & 5.38$\times$3.37 & 0.35$\pm$0.02 & 0.50$\pm$0.04 & 5.93$\pm$0.46  & 1.30$\pm$0.10 & \raisebox{1.5ex}{\alphlc=2.10 $\pm$ 0.09}\\
    \hline
   \end{tabular} 
 \end{minipage}}
\end{table*}
%%%%%%%%%%%%%%%%%%%%%%%%%%%%%%%%%%%%%%%%%%%%%%%%%%%%%%%%%%%%%%%%%%%%%%%%%%%%%%%%%%%%%%%%%
%%%%%%%%%%%%%%%%%%%%%%%%%%%%%%%%%%%%%%%%%%%%%%%%%%%%%%%%%%%%%%%%%%%%%%%%%%%%%%%%%%%%%%%%%
\begin{figure}
\centering
 \includegraphics[bb= 25 15 462 483,clip,angle=0, scale=0.53]{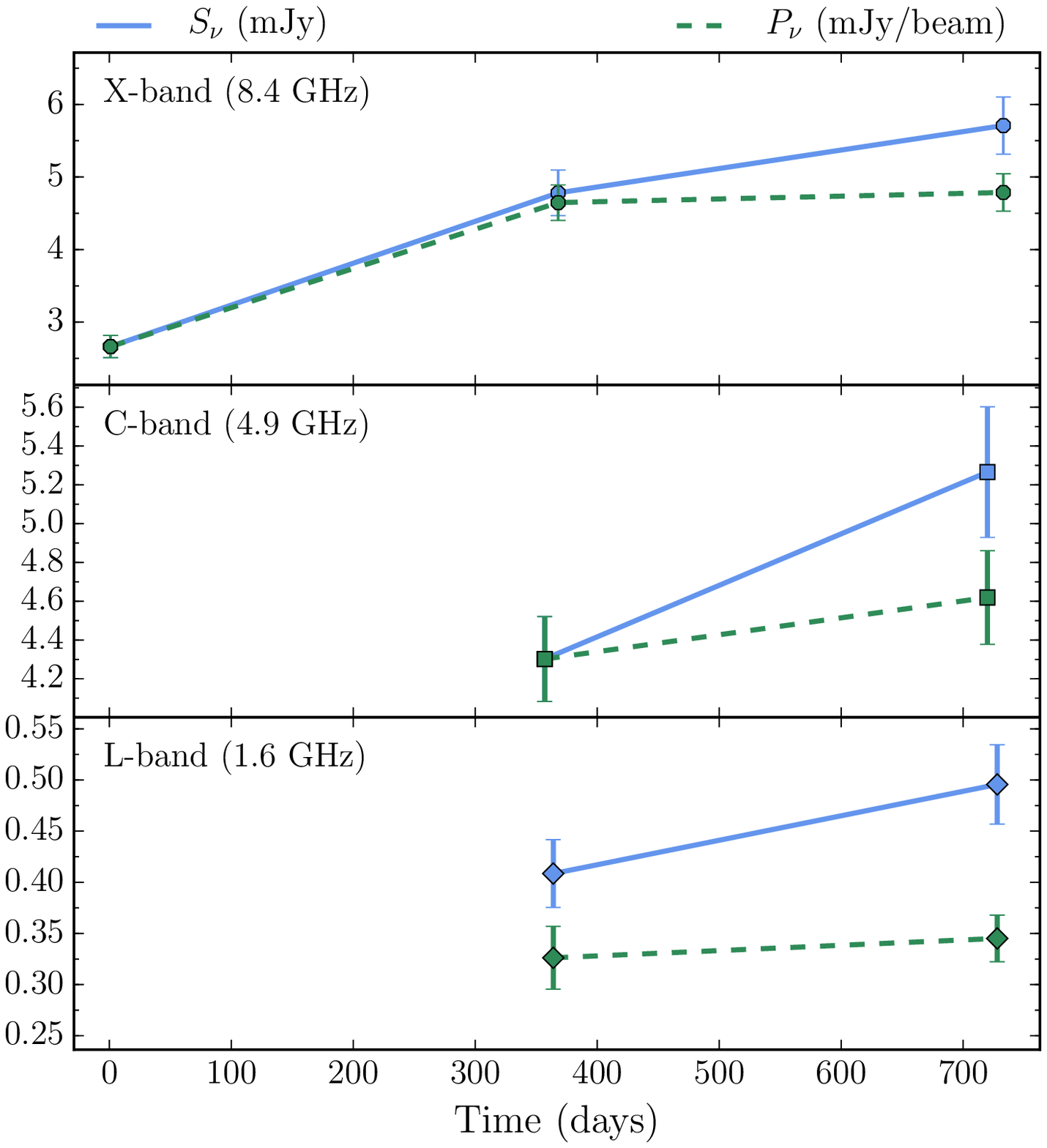} 
 \caption{Evolution of the radio flux density and peak intensity of the nucleus (core plus jet-like  
 structure) at three frequency bands. The error bars represent $\pm1\sigma$ values.}\label{fig:flux_evol}
\end{figure}
%%%%%%%%%%%%%%%%%%%%%%%%%%%%%%%%%%%%%%%%%%%%%%%%%%%%%%%%%%%%%%%%%%%%%%%%%%%%%%%%%%%%%%%%%%%

At 8.4\,GHz there is a significant increase in the flux density from 2012 to 2013. We can infer this from 
both the high-resolution images at $0.98\times0.59$\,mas$^2$, PA$=$2.95\degr{} (Table \ref{tab:xuni}), and 
the lower-resolution images at $7\times6$\,mas$^2$ (Table \ref{tab:lcx_matched}). From 2013 to 2014 there is 
an apparent increase in the total flux density. We note however that adding the emission from the core and 
the jet in the high-resolution images results in similar fluxes within the uncertainties in 2013 and 2014.
Most of the increase in flux we see at lower-resolution is likely due to extended flux unrelated to the jet,
which is recovered by the larger beam. At 4.9 and 1.6\,GHz there is a significant increase in the flux 
densities of about 20 per cent (see Fig. \ref{fig:flux_evol}). Such an increment can also be attributed to 
the presence of extended emission. The peak intensities are consistent with no variation from 2013 to 2014,
and seem to be dominated mostly by the evolution of the core itself.

At 8.4 and at 1.6\,GHz it is clear that the deconvolved sizes are increasing. We attribute 
this to the expansion of the emitting region within the beam. From the high-resolution 8.4\,GHz EVN images
(Fig. \ref{fig:core_jet}), we know there is an emerging jet-like component departing from the core, and we 
associate its appearance to the increase in flux (at 8.4\,GHz) and size of the emitting region measured in 
the lower-resolution images.

The spectral indices remain quite constant (within the uncertainties) from 2013 to 2014, 
being flat between 4.9 and 8.4\,GHz (\alphcx{}) and inverted between 1.6 and 4.9\,GHz
(\alphlc{}), pinpointing to some absorption mechanism. We also note that the luminosities 
($L_{\nu}$) and brightness temperatures (\tb{}) are typical of synchrotron emitting 
sources (Table \ref{tab:lcx_matched}).

As we see in \S \ref{sec:transients}, the large convolving beam in the images we analysed 
in this section, might also include contamination from transient sources.
However these sources would represent only a few percent of the total flux, and
thus their contribution is taken into account within the uncertainties.

\subsection{IC\,883 at high frequencies}\label{sec:spec_vlaemer}

Further insight into the nature of the nuclear region (component A) can be obtained when comparing 
our previously reported e-MERLIN image at 6.9\,GHz \citepalias{rocc12a} with the image from VLA 
A-configuration observations at 32.5\,GHz (see \S \ref{sec_vla}). The combined VLA dataset from June 
2011 was observed somewhat close in time to our e-MERLIN observations (March 2011), and their comparison 
can thus provide us with a rough estimate for the 6.9 to 32.5\,GHz spectral index of the nuclear region.

\begin{figure}
\centering
 \includegraphics[angle=0,scale=0.4]{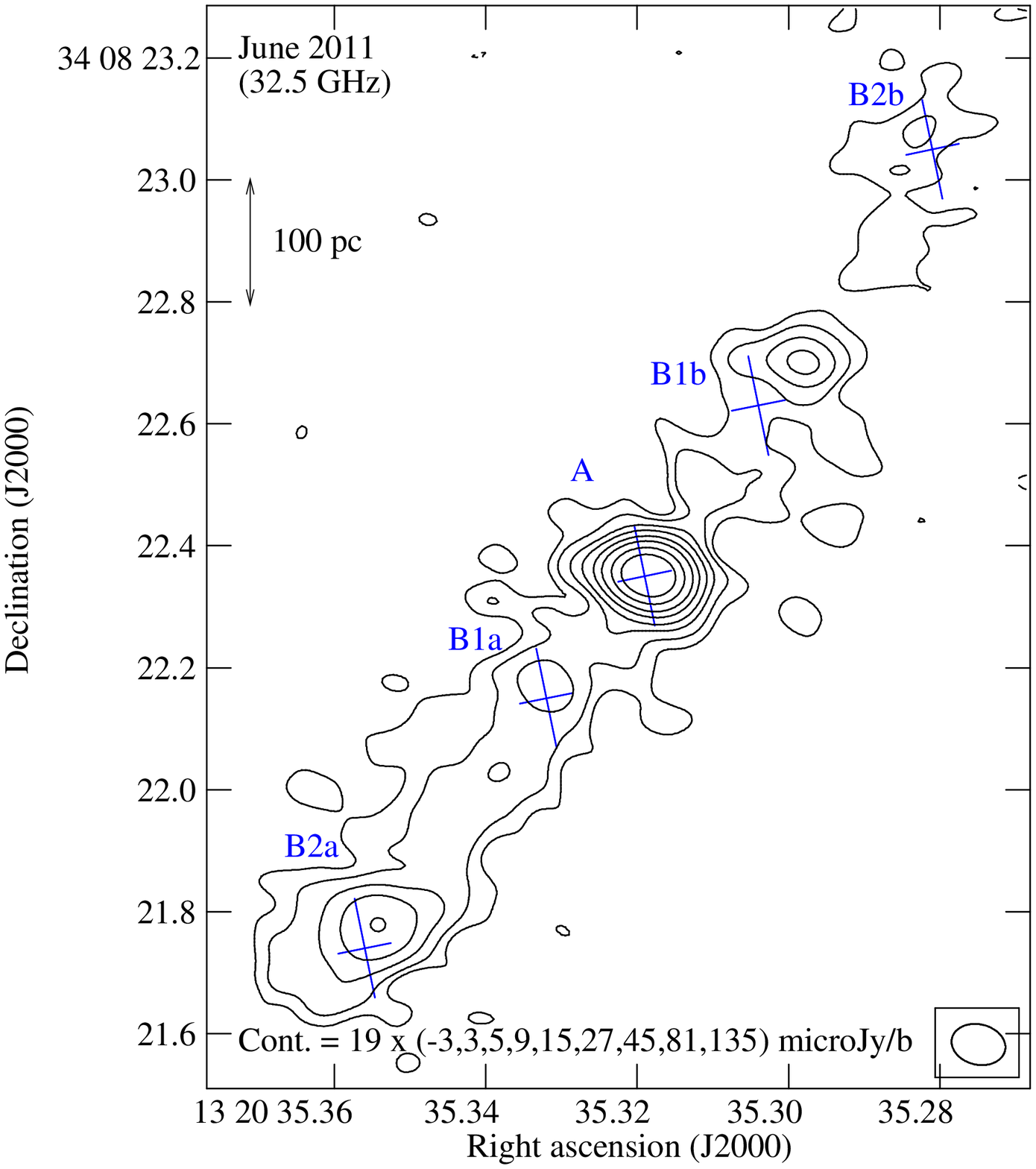}
 \caption{Contour map of the 32.5\,GHz VLA image at a resolution of 89$\times$66\,mas at PA$=$77\degr{}. 
The position of the e-MERLIN detected components \citepalias{rocc12a} are indicated with crosses of the 
size of the convolving beam in the e-MERLIN image (165$\times$88\,mas at PA$=-11.6$\degr{}),
and identified with their corresponding labels.}\label{fig:vlaka}
\end{figure}

We show the contour map of the natural weighted VLA image in Fig. \ref{fig:vlaka}. The image
has a convolving beam of 89$\times$66\,mas at PA$=$77\degr{} and an r.m.s.$=19$\mujyb{}. We recover 
a structure similar to that obtained with e-MERLIN \citepalias[figure 1 
in][]{rocc12a}. We note that from the different knots of emission, only component A has a 32.5\,GHz 
counterpart at the same position. The difference in the rest of the components might arise from the 
fact that at $\nu\ga30$\,GHz we are tracing thermal radiation (e.g., from \hii{} regions), whereas at 
lower frequencies it is the non-thermal radio emission from SNe, SN remnants and even an AGN, that 
dominate the global emission \citep{condon92}. In fact, we obtain brightness temperatures $<10^3$\,K 
for the 32.5\,GHz emission blobs closest to B1a, B1b, B2a and B2b, which agrees with a thermal origin.

The e-MERLIN components B2a, B1a, A and B1b we reported in \citetalias{rocc12a} are at exactly 
the same position as the $^{12}$CO 2--1 clumps reported by \citet{zauderer16}, labelled 
as C1, C2, C3 and C4, respectively. In their study they determined that component C3 
(i.e., component A in our radio observations) is the dynamical centre of the molecular 
ring which shapes the main body of IC\,883. It is in this component that the core-jet
structure is embedded.

%%%%%%%%%%%%%%%%%%%%%%%%%%%%%%%%%%%%%%%%%%%%%%%%%%%%%%%%%%%%%%%%%%%%%%%%%%%%%%%%%%%%%%%%%%%%
%%%%%%%%%%%%%%%%%%%%%%%%%%%%%%%%%%%%%%%%%%%%%%%%%%%%%%%%%%%%%%%%%%%%%%%%%%%%%%%%%%%%%%%%%%%%

To allow a proper comparison of the VLA and the e-MERLIN observations, we have convolved the 
VLA image with the e-MERLIN beam: 165$\times$88\,mas at PA$=-11.6$\degr{}. In the VLA map, at the
position of component A we measure a peak intensity of $4.86\pm0.26$\,\mjyb{} and a flux density
of $5.14\pm0.31$\,mJy. These values are very similar to those reported for component A in 
\citetalias{rocc12a} based on the e-MERLIN map. Hence, we obtain a two-point spectral index 
between 6.9 and 32.5\,GHz of $-0.03\pm0.05$. Assuming that at 32.5\,GHz the milli-arcsec core 
(component A1) will also be the dominant radio emitter from component A, as noted at 6.9\,GHz 
with e-MERLIN, we infer that the core has also a very flat spectral index all the way from 8.4 
to 32.5\,GHz. 

Using a spectral index of $-0.03\pm0.05$ for the core and the flux density at 8.4\,GHz from 
Table \ref{tab:lcx_matched}, we calculate a 32.5\,GHz flux density of $4.60\pm0.44$\,mJy in 
2013, and $5.50\pm0.54$\,mJy in 2014.

\subsection{Radio SED of the nucleus}\label{sec:sed}

In Table \ref{tab:xuni} we reported the 8.4\,GHz flux densities measured in 2014 for both
the core and the jet. The core is approximately three times as luminous as the jet, and thus
its emission will be the major contributor to the total nuclear flux measured at $7\times6$\,mas$^2$.

In 2012 and 2013 the jet component has not been completely ejected, and we expect that its contribution 
to the total flux is smaller than in epoch 2014. The core will be the dominant emitter in 2012. Since 
we have only measurements at one frequency in 2012, we resort to considering the 2013 epoch as closest
to quiescence, and use it to investigate the nature of the nucleus.

The high luminosities ($L_{\nu}$) and brightness temperatures (\tb{}) estimated for the core 
and the jet (Tables \ref{tab:xuni} and \ref{tab:lcx_matched}), are indicative of 
a synchrotron origin for the radio emission at all frequencies. We note that the spectrum is highly 
inverted and it shows a turnover at low frequencies. Free-free absorption (FFA) and synchrotron 
self-absorption (SSA) are the most commonly assumed mechanisms responsible for the turnover. We thus
attempted to fit the radio SED with the available data in 2013 considering both pure SSA and pure FFA. 
Owing to the scarcity of data, we focused our efforts in estimating the turnover frequency ($\nu_{t}$) 
in the SED. 

%%%%%%%%%%%%%%%%%%%%%%%%%%%%%%%%%%%%%%%%%%%%%%%%%%%%%%%%%%%%%%%%%%%%%%%%%%%%%%%%%%%%%%%%%%%%%%%%%%%%%%%%%%%%%%%%%% 
\begin{figure}
\centering
 \includegraphics[bb= 0 15 468 468,clip,angle=0, scale=0.53]{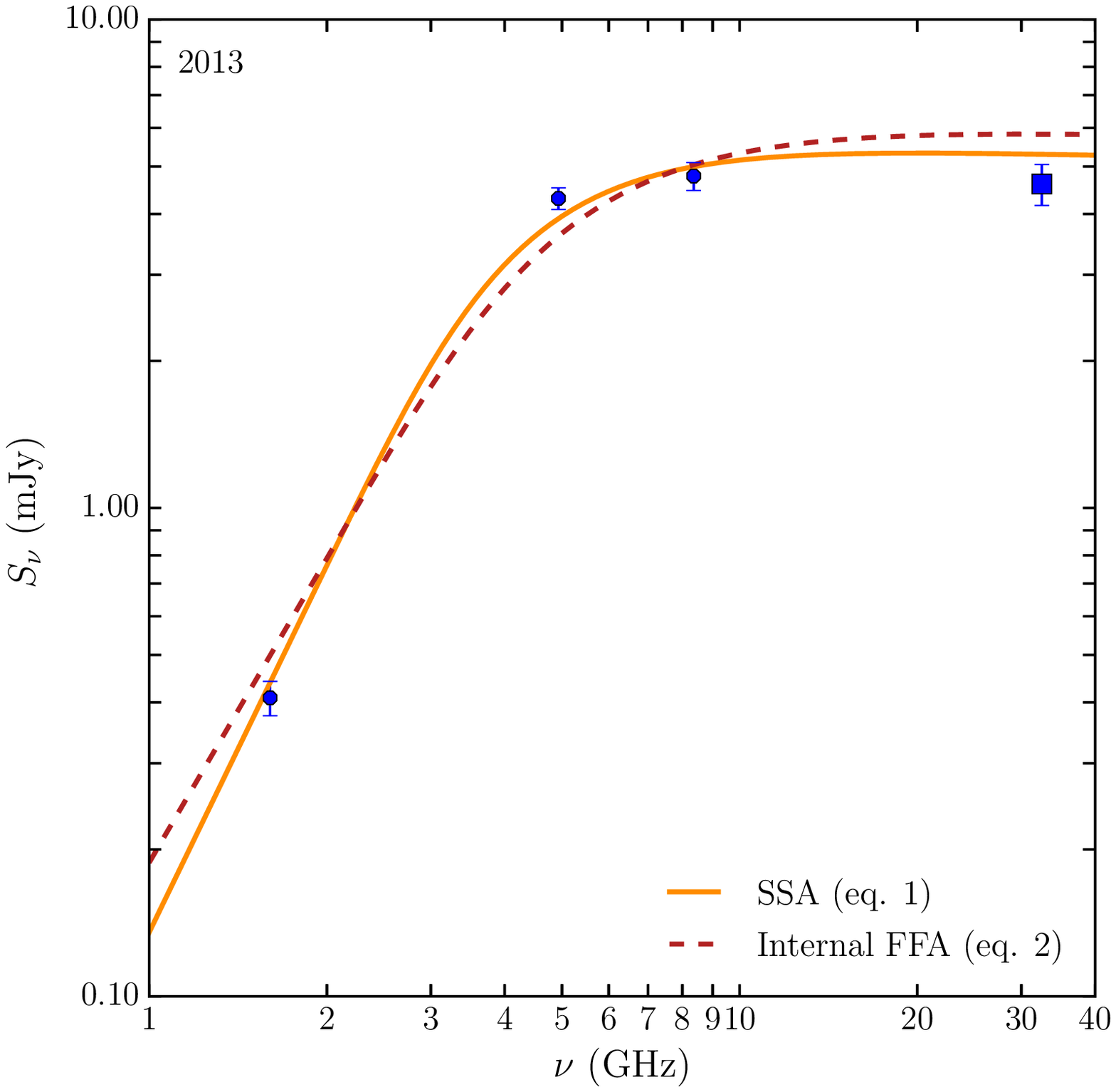} 
 \caption{Radio spectral energy distribution of the core from contemporaneous measurements at three 
different frequencies in 2013 (circles) and the estimated flux density at 32.5\,GHz (square). 
Two different fits assuming SSA (solid line) and FFA (dashed-line) have been applied to the data 
as described in the legend inside the plot and in the text. The error bars are at a $1\sigma$ 
level.}\label{fig:extended_rsed}
\end{figure}
%%%%%%%%%%%%%%%%%%%%%%%%%%%%%%%%%%%%%%%%%%%%%%%%%%%%%%%%%%%%%%%%%%%%%%%%%%%%%%%%%%%%%%%%%%%%%%%%%%%%%%%%%%%%%%%%%% 

\subsubsection{Synchrotron self-absorption}\label{sec:ssa}

The spectral index between 1.6 and 4.9\,GHz (\alphlc{} in Table \ref{tab:lcx_matched}) is 
very close to the spectral index expected from SSA in a homogeneous source ($\alpha=2.5$). 

To construct the SED we can use a bending power-law, which produces a smooth transition
between the optically-thick and optically-thin parts of the SED. Another possibility is
to use a broken power-law, but this approach does not have an empirical basis.

For the bending power-law, we consider $\alpha_1=2.5$ for the optically-thick part. 
We do not have information at frequencies where the spectrum becomes completely optically 
thin, but we consider $\alpha_2=-0.03$ to describe it. Following \citet{callingham15}, the
flux density at the frequency $\nu$ is described by

\begin{equation}\label{eq:bepl}
S_{\nu} =S_0\left( \frac{\nu}{\nu_t} \right)^{\alpha_1} \left( 1 - \rmn{e}^{-\left( \frac{\nu}{\nu_t} \right)^{\alpha_2-\alpha_1} }\right)
\end{equation}
where $S_0$ is a constant. 

We performed a non-linear least-squares minimisation to fit the bending power-law to the data
and solved for $S_0$ and $\nu_t$. We fitted our EVN data which covers the evolution of the 
core between 1.6 and 8.4\,GHz, and included the expected flux density at 32.5\,GHz, which 
provides us with an additional degree of freedom (dof). The goodness of the fit given by 
$\chi^{2}$ resulted in a value of 3.5 for 2 dof, and a turnover at $4.44_{-0.62}^{+0.72}$\,GHz. 
The resulting SED is shown as a solid line in Fig. \ref{fig:extended_rsed}.

We also performed the fit using $\alpha_2=-0.7$, which is a typical spectral index for the
optically-thin part of the SED \citep{orienti15}. However, this leads to $\chi^{2}$ values
$\geq20$, consequently with a p-value $\approx0$. We acknowledge that the scarcity of data 
reduces the reliability of the fits and allows the questioning of the assumptions we have 
made, although these seem to be quite fair.

As the behaviour of the core is similar for the fluxes and spectral indices in both 2013 and 
2014, for completeness, we have repeated the above process for the values measured in 2014. 
This led to a very similar turnover frequency of $4.42_{-0.59}^{+0.68}$\,GHz.

\subsubsection{Free-free absorption}

There is ongoing star formation activity in the nuclear region of IC\,883 (see \S \ref{sec:transients} 
and \ref{sec:ongoing_sf}), hence internal FFA could be the mechanism responsible for producing the 
inverted spectrum we observe. We assume that the flux density follows:
\begin{equation}\label{eq:intffa}
S_{\nu} =S_0 \nu^{\alpha_2} \left( \frac{\nu}{\nu_t} \right)^{2.1} \left( 1 - \rmn{e}^{-\left( \frac{\nu}{\nu_t} \right)^{-2.1} }\right)
\end{equation}
\citep[as described by][]{callingham15} where we used the same notation as in \S \ref{sec:ssa}. 
Since the spectral index in the optically-thin part of the spectrum is not known, we proceeded
with $\alpha_2=-0.03$ as in \S \ref{sec:ssa}. The SED obtained this way is shown in Fig. 
\ref{fig:extended_rsed}. From visual inspection we infer that the fit using FFA is slightly 
worse than that obtained using SSA. In fact, we obtain a larger $\chi^{2}$ value, 13 for 2 dof 
and assuming $\alpha_2=-0.03$, and 49.5 for 2 dof and assuming $\alpha_2=-0.7$. The turnover 
frequency in this case is 5.4\,GHz. Similar values are obtained when fitting the 2014 measurements.

Three or four data points are not sufficient to assess the reliability of either SSA or FFA, and 
although we cannot neglect the influence of FFA, our fit favours SSA, and thus this is the absorption 
mechanism we adopt hereafter.

\subsection{Transient radio sources in IC\,883}\label{sec:transients}

Each of our EVN observations achieved an r.m.s.$>15$\,\mujyb{} (see Table \ref{tab:lcx_matched}) 
which is equivalent to a $<5.6\sigma$ detection of a radio supernova with luminosity 
$\sim1\times10^{27}$\lunits{} (e.g., SN\,1993J).

To ease the detection of radio transients in our observations, we created deep images 
following two paths in {\sc aips}: (i) combining epochs in the image plane; and
(ii) imaging the concatenated {\it uv}-data at each frequency. The former path gave unreliable 
results, as it enhanced artifacts present in individual images. The latter path led to a
decrease in the r.m.s. We additionally attempted phase-only self-calibration to further improve 
image quality. However, this process failed to provide good solutions given that our target
sources are faint.

\begin{figure}
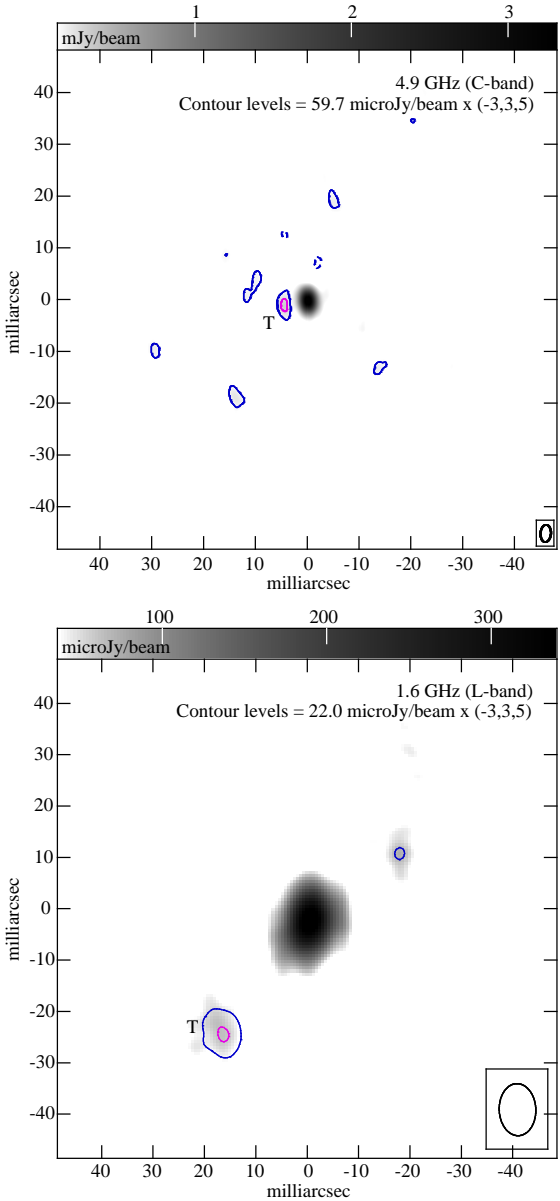

\centering
 \includegraphics[scale=0.4]{fig5u.eps}
 \includegraphics[scale=0.4]{fig5l.eps}
 \caption{Grey-scale maps of the 4.9 (top panel) and 1.6\,GHz deep images, with
overlaid contours of the images where the nuclear emission has been subtracted. 
Putative transient sources are labelled with a {\bf T}.}\label{fig:deep}
\end{figure}

Using the concatenated {\it uv}-data we made naturally-weighted deep images, which are best to test 
detection feasibility. These images are shown in grey-scale maps in Fig. \ref{fig:deep} (excluding 
the 8.4\,GHz frequency band) and their characteristics are listed in Table \ref{tab:deep}.

%%%%%%%%%%%%%%%%%%%%%%%%%%%%%%%%%%%%%%%%%%%%%%%%%%%%%%%%%%%%%%%%%%%%%%%%%%%%%%%%%%%%%%%%%
\begin{table}
\centering
\caption{\protect{Characteristics of naturally-weighted deep images.}} \label{tab:deep}
\begin{tabular}{ccc} \hline
Frequency & FWHM, PA       & r.m.s.   \\
(GHz)&  (mas$^2$, \degr{}) & (\mujyb) \\
    \hline
8.4 & $1.45 \times 0.77$, $0.1$  & 48.6 \\
4.9 & $3.34 \times 2.00$, $-4.1$ & 42.8 \\
1.6 & $10.14 \times 7.12$, $2.4$  & 19.1 \\
    \hline
   \end{tabular} 
\end{table}
%%%%%%%%%%%%%%%%%%%%%%%%%%%%%%%%%%%%%%%%%%%%%%%%%%%%%%%%%%%%%%%%%%%%%%%%%%%%%%%%%%%%%%%%%

%%%%%%%%%%%%%%%%%%%%%%%%%%%%%%%%%%%%%%%%%%%%%%%%%%%%%%%%%%%%%%%%%%%%%%%%%%%%%%%%%%%%%%%%%%%%
%%%%%%%%%%%%%%%%%%%%%%%%%%%%%%%%%%%%%%%%%%%%%%%%%%%%%%%%%%%%%%%%%%%%%%%%%%%%%%%%%%%%%%%%%%%%%%%%

We created new images at the same spatial resolution as the deep images, but restricting 
the cleaning to the very central region, which is dominated by the core and the jet. We then
subtracted these images from their corresponding concatenated datasets. This resulted in 
{\it uv}-datasets that are core-jet emission free, whose images are shown as contours 
overlaid on top of the grey-scale maps in Fig. \ref{fig:deep}. The contours represent the 
residuals, and we identify two putative transient sources with a S/N$>5$ (labelled with a 
{\bf T}), and thus correspond to robust detections. We carefully checked that 
these detected sources do not lie on top of sidelobes. There is no obvious emission left after 
the subtraction of the nuclear emission at 8.4\,GHz. We note as well that previously known 
nuclear components \citepalias[sources A2, A3 and A4 from ][]{rocc12a} were not detected, hence 
confirming their transient nature.

To further test the reliability of the {\bf T} sources, we produced natural weighted maps of 
the individual epochs at the resolution indicated in Table \ref{tab:deep} according to the 
corresponding frequency. The image fidelity is slightly worse than in the deep images as there 
are fewer visibilities, however we find that the {\bf T} source at 4.9\,GHz is detected only in 
epoch 2013, and the 1.6\,GHz {\bf T} source appears only in epoch 2014. The 4.9\,GHz {\bf T} 
source is still marginally detected in a uniformly-weighted map, whilst the 1.6\,GHz is washed 
out when using that weighting scheme. We therefore report only one reliable transient source in 
epoch 2013 at 4.9\,GHz, with a luminosity of $(4.9\pm1.2)\times10^{27}$\lunits{}, based on the 
4.9\,GHz deep radio image. Such a luminosity is comparable to that of bright radio SNe at their 
maximum light.

\subsection{{\it XMM-Newton} and {\it Chandra} results}\label{sec:xray}

\begin{figure}
\centering
 \includegraphics[angle=270, scale=0.35]{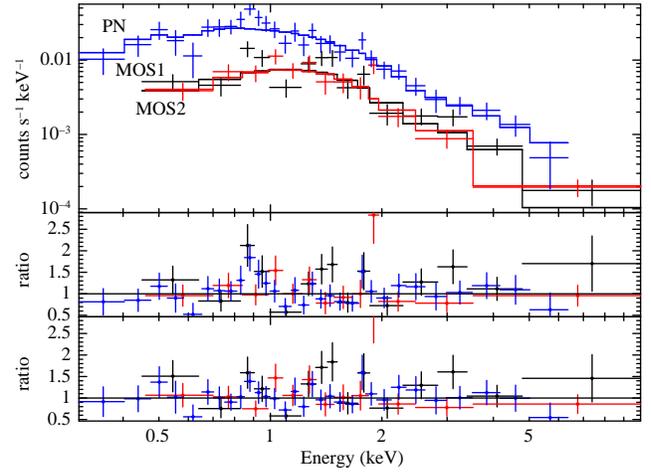} 
 \caption{{\it XMM-Newton} spectrum of IC\,883. Middle and lower panels correspond to the data 
to model residuals, {\sf tbabs}*{\sf powerlaw} and {\sf tbabs}*({\sf powerlaw} $+$ {\sf apec}), 
respectively.}\label{fig:xmmspec}
\end{figure}

The top panel of Fig. \ref{fig:xmmspec} shows the three {\it XMM-Newton} spectra 
in the 0.2--10\,keV range: the EPIC/PN spectrum is shown in blue and the MOS 1 and MOS 
2 spectra in black and red respectively. For all fits we used the {\sc xspec} package 
and fit the three spectra simultaneously with all parameters tied.

We first tried a simple fit with a power-law model ({\sf powerlaw}) obscured by neutral material 
using the Tuebingen-Boulder absorption model for X-ray absorption by the gas, molecular 
and dust phases of the interstellar medium (ISM) \citep[{\sf tbabs};][]{wilms00}, with a 
hydrogen column density fixed at the Galactic value in the direction of the source, 
$N_{\rmn{H}}=10^{20}$\,cm$^{-2}$. This model gives a poor fit with $\chi^2=130$ for 
57\,dof and obvious residuals peaking sharply at 0.8\,keV, and overestimates
the flux at low and high energies. Freeing the column density and maintaining the {\sf powerlaw} 
slope and normalisation free improves the fit significantly to $\chi^2=74.5$ for 56\,dof. This 
model is plotted as a solid line in the top panel of Fig. \ref{fig:xmmspec} and the ratio between 
the data and the model is plotted in the middle panel.  The best fitting values
of the parameters are $N_{\rmn{H}}=(1.8 \pm 0.3)\times 10^{21}$\,cm$^{-2}$, 
photon index $\Gamma = 2.1\pm0.1$, and normalisation $(3.7 \pm 0.4)\times 10^{-5}$\,\phunits{}
at 1\,keV. These values result in an unabsorbed 2--10 keV flux of $8.11\times10^{-14}$\funits{} 
and a 2--10\,keV luminosity $L_{2-10\rmn{keV}} = 9.75\times 10^{40}$\rlunits{}.

\begin{figure}
\centering
 \includegraphics[scale=0.4]{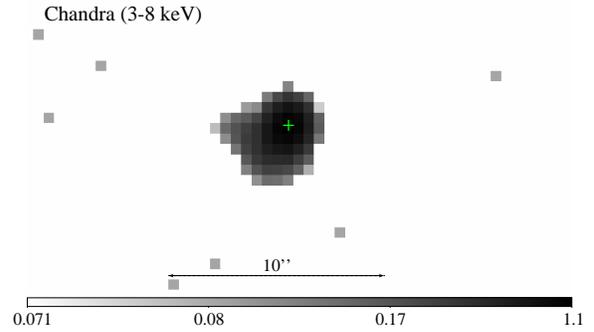}
 \caption{Hard X-rays {\it Chandra} image centred on IC\,883. The units are counts smoothed 
 with a Gaussian kernel of radius 3 pixels. The cross represents the position of the radio 
 core.}\label{fig:chandraim}
\end{figure}

\begin{figure}
\centering
\includegraphics[scale=0.35,angle=270]{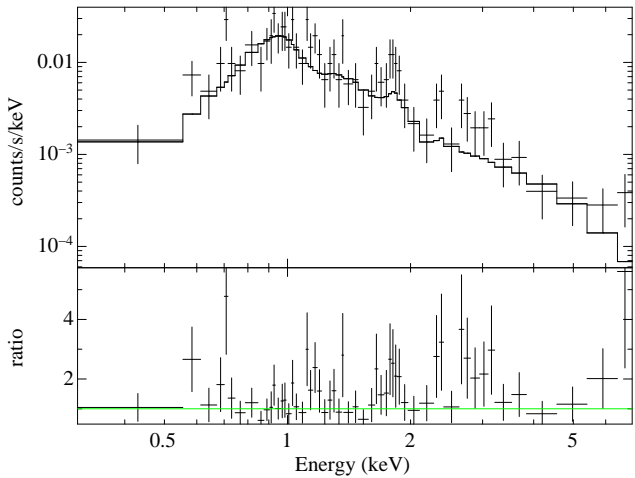}
  \caption{{\it Chandra}/ACIS spectrum of IC\,883 fit with a model that includes an
absorbed power-law and collisionally ionised thermal component. The bottom panel
shows the ratio between the model and the data.}
\label{fig:Chandra_spec}
\end{figure}

The higher angular resolution {\it Chandra} images of IC\,883 show a spatially extended emission 
region around a central point source (Fig. \ref{fig:chandraim}). This extended emission is inevitably 
included in the extraction region of the PN and MOS source spectra. This emission probably corresponds 
to hot gas in the galaxy and we account for it by adding an optically-thin thermal emission component,
using the Astrophysical Plasma Emission Code \citep[{\sf apec};][]{smith01}. For the {\it XMM-Newton} 
spectrum, fitting the model {\sf tbabs}*({\sf powerlaw} $+$ {\sf apec}) with free column density, 
{\sf powerlaw} slope and normalisation, and {\sf apec} temperature and
normalisation, while keeping the abundance fixed to the Solar value and the
redshift at the systemic value $z=0.023$, gives a slightly better fit with
$\chi^2=63.6$ for 54\,dof. The hot gas component is always sub-dominant but
makes its largest contribution between 0.7 and 1\,keV removing a small peak
in the residuals at these energies. The ratio of the data to this model is
shown in the bottom panel in Fig. \ref{fig:xmmspec}, where the peak at 0.85
keV, most clearly seen in the PN (blue points) is better modelled. The best
fitting parameters are $N_{\rmn{H}}=(1.57 \pm 0.35)\times10^{21}$cm$^{-2}$, 
$\Gamma = 1.85\pm0.14$, {\sf powerlaw} normalisation $=(2.8\pm0.4)\times 10^{-5}$\phunits{} 
at 1 keV, {\sf apec} temperature $=0.86\pm0.14$\,keV and {\sf apec} normalisation 
$(4.6\pm 1.4)\times 10^{-6}$\phunits{} at 1\,keV. These values result in
an unabsorbed 2--10 keV flux of the {\sf powerlaw} only of $9.0^{+1.0}_{-5.2}\times10^{-14}$\,\funits{} 
and $L_{2-10\rmn{keV}} = 1.07\times 10^{41}$\rlunits{}, similar to the value reported
by \citet{modica12}.
{\it Chandra} data were binned to 1 count per bin, and we used Cash statistics to
fit the ACIS-S spectrum, resulting in $C=72.2$ for 147\,dof. Using a model similar 
to the one applied to {\it XMM-Newton} data we obtain a column density of 
$N_{\rmn{H}}=1.8^{+1.9}_{-1.4}\times10^{21}$\,cm$^{-2}$, 
a photon index of $\Gamma=1.80_{-0.75}^{+0.78}$, and a temperature of the {\sf apec} of
$kT=1.00^{+0.28}_{-0.30}$\,keV. All these parameters are consistent with the values
obtained with {\it XMM-Newton}/EPIC. The {\it Chandra} spectrum is shown in 
Fig.\,\ref{fig:Chandra_spec}. The observed 2--10\,keV flux  is
$(6\pm3)\times10^{-14}$\,\funits{}.

%%%%%%%%%%%%%%%%%%%%%%%%%%%%%%%%%%%%%%%%%%%%%%%%%%%%%%%%%%%%%%%%%%%%%%%%%%%%%%%%%%%%%%%%%%
%%%%%%%%%%%%%%%%%%%%%%%%%%%%%%%%%%%%%%%%%%%%%%%%%%%%%%%%%%%%%%%%%%%%%%%%%%%%%%%%%%%%%%%%%%%

%%%%%%%%%%%%%%%%%%%%%%%%%%%%%%%%%%%%%%%%%%%%%%%%%%%%%%%%%%%%%%%%%%%%%%%%%%%%%%%%%%%%%%%%%%%%
%%%%%%%%%%%%%%%%%%%%%%%%%%%%%%%%%%%%%%%%%%%%%%%%%%%%%%%%%%%%%%%%%%%%%%%%%%%%%%%%%%%%%%%%%%%

%%%%%%%%%%%%%%%%%%%%%%%%%%%%%%%%%%%%%%%%%%%%%%%%%%%%%%%%%%%%%%%%%%%%%%%%%%%%%%%%%%%%%%%%%%
%%%%%%%%%%%%%%%%%%%%%%%%%%%%%%%%%%%%%%%%%%%%%%%%%%%%%%%%%%%%%%%%%%%%%%%%%%%%%%%%%%%%%%%%%%

\section{Discussion}\label{sec:discussion}

The high-resolution radio data we have presented in this paper provide unambiguous evidence of
the existence of a jet in IC\,883, and thus, for the presence of an AGN in the nucleus. On the 
other hand, our analysis of the available X-ray data agrees both with star formation and AGN
activity in IC\,883. In this section we discuss the observational evidence in order 
to understand the ongoing physical processes in the core of IC\,883.

\subsection{Origin of the X-ray emission}

The origin of the X-ray emission is difficult to determine. The 
0.2--10\,keV spectrum contains an optically-thin thermal
component and a {\sf powerlaw} with slope $\Gamma =1.8$. The former can be
produced by a collisionally ionised plasma in a star forming region
while the second can be produced by a related population of high-mass
X-ray binaries. In this case, the entire X-ray emission can be
ascribed to star formation. Alternatively, a low luminosity Seyfert
spectrum could produce the powerlaw emission and photo-ionise its
surroundings to produce most of the emission below 2\,keV, which is also
consistent with the optical LINER classification of this source.

Taking into account the background-corrected {\it Chandra} counts in the 2--8\,keV ($H$) and
0.5--2\,keV ($S$) bands, the hardness ratio ($HR=(H-S)/(H+S)$) for IC\,883 is
$HR=-0.56$. This value is below the threshold set by \citeauthor{iwasawa11} for AGN ($HR>-0.3$),
and is consistent with the average of the GOALS sample \citep[$HR=-0.56$,][]{iwasawa11}.
However, a slope of $\Gamma =1.8$, as measured here, is also consistent with a low-accretion 
rate AGN \citep[e.g.,][]{fanali13}.

The X-ray to star formation rate (SFR) correlation of \citet{ranalli03}, considering the SFR of 
IC\,883 \citepalias[185\,\msun\,yr$^{-1}$,][]{rocc12a}, shows that the expected 2--10\,keV
luminosity solely due to star formation would be $\sim 9.45\times10^{41}$\,\rlunits{}, which is
10 times higher than the measured X-ray luminosity of the source. This discrepancy can be
related to a high obscuration of the star forming region itself. Furthermore, the high
X-ray flux expected would suggest that most of the observed X-ray emission arises from
the star formation.

\subsection{A new jet component}

Our EVN observations on 4 November 2014 clearly show a resolved core-jet structure in IC\,883.
The jet was likely ejected some time between 15 May 2011 and 2 November 2012 hence moving away 
from the core with an apparent speed of 0.6--1.0\,$c$ (see Section \ref{sec:jet}). Such subluminal 
proper motion is comparable to the apparent velocities of jets detected in LLAGNs 
\citep[e.g.,][]{nagar02b,mezcua14,argo15}.

We note that the appearance of the jet component was not associated with a strong radio flare 
in IC\,883 at least since 2004, time of the first VLBI observations at frequencies 
higher than 5\,GHz reported in \citetalias{rocc12a} where the core is clearly detected, though
even before that, \citet{lonsdale93} reported an upper limit at 1.49\,GHz from global VLBI 
observations made in 1991. The jet is detected thanks to the high-resolution and high-sensitivity
observations that were made toward IC\,883 for the first time. Similar observations are needed
to unveil parsec-scale jet components in other AGNs as suggested by \citet{mezcua14}. We are probably 
witnessing recurrent AGN activity in IC\,883 rather than its onset, unlike in the case of NGC\,660 where 
the radio brightness increased by a factor of $>1200$, accompanying the appearance of a new compact 
source and the production of a jet \citep{argo15}.

Given the nuclear flux density level in 2011 \citepalias[4.38$\pm$0.28\,mJy,][]{rocc12a} and its 
decrease by 2012 (see Table \ref{tab:xuni}), we can infer there was probably a previous component 
similar to the one we clearly detect in 2014. In \citetalias{rocc12a} we also reported variability 
in the flux density of the nucleus at 5\,GHz (labelled as component A1 in that paper) in the unevenly 
sampled period from 2004 to 2011. IC\,883 could be producing jet components continuously. However, 
the jet components do not seem to live very long, as we do not see larger-scale jets, not even at a 
2\,mas resolution. 

It is thought that a very dense ISM (as expected in the nuclear region of LIRGs) can halt
jets moving through it \citep{bicknell03}. The interaction of the jet with the ISM can be probed
by means of \hi{} absorption observations. However, the possibility exists that \hi{} absorption 
also probes foreground molecular clouds which happen to lie in the line of sight \citep{conway99},
as seems to be the case here (see \S \ref{sec:intro}). In IC\,883, the \hi{} absorption
reported by \citet{clemens04} is displaced by $0.2$\,arcsec ($\approx 100pc$) from component A 
\citep[][]{zauderer16}, and hence it is not probing directly the core-jet region. 

\citet{papadopoulos14} report a deficit of dense gas toward the dynamical centre region 
(which corresponds to the nuclear region as we have seen at radio frequencies). Furthermore, 
\citet{modica12} found evidence of a nuclear stellar wind based on the soft X-ray emission 
and morphology. In fact, the soft X-ray spectrum can be well reproduced by a collisionally 
ionised plasma, which is a typical signature of star formation. We can exclude the presence 
of a strong X-ray jet, since its radio counterpart is not strong. Is the jet clearing its 
way through the surrounding gas? Probably not, as it is more likely that the star formation 
feedback is responsible for the gas dispersal.

What is then hindering the production of large scale radio jets and keeping the central radio 
source confined to very small regions? The reason could simply be that the AGN is
producing weak jet components which are not powerful enough to break through the dense ISM.

\subsection{A Gigahertz Peaked-Spectrum galaxy candidate}

The IC\,883 nucleus has a spectral index between 1.6 and 4.9\,GHz (in 2013 and 2014) very close to 
the spectral index expected from synchrotron self-absorption in a homogeneous source ($\alpha=2.5$),
with a turnover frequency of $4.44_{-0.62}^{+0.72}$\,GHz. This characteristic, together with 
the compactness of the nucleus (at a few mas-scale), are typical features of the so-called 
Gigahertz-Peaked Spectrum (GPS) sources \citep[e.g.,][]{odea91,devries97,snellen09}, which also share 
common properties with LLAGNs \citep{nagar02a}. GPS sources are important as they represent 
the first stages of powerful radio galaxies \citep[e.g.,][and references therein]{odea98, snellen00}.

GPS galaxies have typical luminosities at 5\,GHz above $10^{31}$\,\lunits{} ($10^{24}$\,W\,s$^{-1}$). 
This is about three orders of magnitude higher luminosity than that displayed by IC\,883 (see Table 
\ref{tab:lcx_matched}). However, it is known that current GPS catalogues are biased toward bright 
sources at redshifts $>0.1$. The discovery by \citet{tingay97} of the nearest GPS galaxy at the time 
posed the question whether the low-luminosity, low-redshift GPS sources are hiding among or within LINER 
sources \citep{odea98}. In fact, \citet{tingay03} presented three nearby GPS sources hosted by elliptical 
galaxies which can be thought of as IC\,883 ``cousins'' based on the common properties shared by 
their hosts: {\it (i)} showing strong evidence of interacting/merging activity; {\it (ii)} showing 
optical spectra identifying them as LINERs; {\it (iii)} being low-redshift galaxies ($<$100\,Mpc) 
with luminosities below the average for GPS galaxies \citep[see also][]{hancock09, tingay15}, although 
IC\,883 remains the least luminous one.

It has been suggested that the compactness (on sub-kpc scales) of GPS galaxies is due to interaction with 
very dense surrounding gas, which is responsible for hindering the jet growth \citep[e.g.,][]{bicknell97},
although there is now mounting evidence that their sizes can be explained in terms of their age 
\citep{odea98,snellen09}. There are different methods to determine the age of GPS sources. For instance, 
using the curvature of the synchrotron spectrum in a sample of compact lobe-dominated objects and assuming 
equipartition magnetic fields, \citet{murgia03} found spectral ages between 10$^2$--10$^5$\,yr for such 
objects. \citet{polatidis03} measured the subluminal proper motion of hot spots, and determined kinematic 
ages within the same range. However, \citet{murgia03} noted that when the morphology of the 
GPS source is dominated by strong jets or hot spots, the spectral age will be a lower limit to the 
source age.

In the case of IC\,883 there is no available information about the frequency at which the spectrum breaks 
and becomes steep (i.e., where radiation losses dominate) and thus we cannot determine the spectral age.  
In an attempt to estimate the age of the AGN in IC\,883, we consider the formulation for the
magnetic field and synchrotron time given by \citet{pacholczyk},

{\scriptsize
\begin{equation}\label{eq:Beq}
\left(\frac{B_{\rmn{eq}}}{\mu\rmn{G}}\right) \approx 8.1 \left[ \frac{(1+k)}{\phi} \left(\frac{c_{\mbox{\tiny 12}}}{10^7}\right)
            \left(\frac{R}{1\,\rmn{kpc}} \right)^{-3} \left(\frac{ L_{\rmn{R}}}{10^{39}\,\rlunits{}}\right) \right]^{2/7}
\end{equation}}
\noindent where $\phi$ is the filling factor of fields and particles, $k$ is the ratio of heavy particle 
energy to electron energy, adopting for simplicity the values of 0.5 and 100, respectively
\citep[see e.g.,][]{beck05}. $R$ is the linear size of the radio emitting region, with a value of 
7.3\,pc (15\,mas), as measured from the EVN images at $7\times6$\,mas$^2$. $c_{\mbox{\tiny 12}}$ 
is a function of the upper and lower frequencies considered and their corresponding spectral index,
which in this case is $\sim7.76\times10^6$.

From Table \ref{tab:lcx_matched}, we see that \alphcx{} varied from 0.20 in 2013 to 0.15 in 2014.
This is the turnover region of the radio SED. We have also determined that at higher frequencies, 
the core has a much flatter spectral index ($-0.03\pm0.05$), which holds at least from 8.4 to 
32.5\,GHz (see \S \ref{sec:spec_vlaemer}). Using these as lower and upper frequencies, we calculate 
a radio luminosity of $(1.3\pm0.2)\times10^{39}$\rlunits{} in 2013 and $(1.6\pm0.2)\times10^{39}$\rlunits{} 
in 2014. We thus determine a magnetic field strength of $2.6\pm0.2$\,mG in 2013, and a similar value in 2014 
that differs by only 5 per cent.

\citet{tyulbashev01} obtained mean magnetic field strengths of the order of 1--10\,mG for a sample 
of GPS sources,  in good agreement with the $B_{\rmn{eq}}$ values compiled by \citet{odea98} from 
the literature. The value we estimated for IC\,883 is consistent with the values reported by those  
authors. It is remarkable that even though IC\,883 is also an advanced merger, its magnetic field 
strength exceeds by at least an order of magnitude that of other sources belonging to the same class 
\citep{momjian03,drzazga11,rocc12b}. 

We can now calculate the lifetime of the electrons with minimum energy moving in such a magnetic 
field and at a frequency $\nu_{\rmn{c}}$ (considered here as 8.4\,GHz) as:
\begin{equation}\label{eq:tsyn}
\left( \frac{\tau_{\rmn{syn}}}{\rmn{Myr}}\right) \approx
1.06\times 10^3\left[\left(\frac{\nu_{\rmn{c}}}{\rmn{GHz}}\right)\left(\frac{B_{\rmn{eq}}}{\mu
\rmn{G}}\right)^3\right]^{-1/2}
\end{equation}
\citep{pacholczyk}, resulting in an age of $(2.8\pm0.3)\times10^3$\,yr, which is at the lower end of the 
typical age range for GPS galaxies. An estimate of the electron lifetime according to  \citet{vanderlaan69} 
leads to basically the same result ($t_{\rmn{e}}=3.1\times10^3$\,yr). In the following, we adopt an age of 
$\sim3\times10^3$\,yr for the IC\,883 core.

\subsection{Ongoing star formation in the system}\label{sec:ongoing_sf}

In \citetalias{rocc12a} we estimated a core-collapse SN (CCSN) rate of $1.1_{-0.6}^{+1.3}$\,yr$^{-1}$
and a SFR$=$185\,\msun\,yr$^{-1}$ for IC\,883. Such high rates make IC\,883 a prime target
for CCSN searches using high-spatial resolution radio and near-IR observations.

We have concatenated all the EVN epochs to produce deep images of IC\,883 at different frequencies
to facilitate the SN search. We detected a transient source at 4.9\,GHz in 2013 with a luminosity
comparable to that of luminous radio SNe and give further evidence of the ongoing star 
formation activity in the nuclear regions of IC\,883. 

If the transient is indeed a SN, it seems to evolve fast, as we do not detect it in consecutive 
yearly epochs and it appears at only one frequency at a time. Another 
possibility is that this is a slowly evolving SN and it appears to be caught close to its maximum 
light, and had already fallen below the detection limit in a timespan of a year.

On account of IC\,883's distance, the discovery of transients in the radio is biased to only the 
brightest SNe. SNe 2010cu and 2011hi exploded in the circumnuclear regions of IC\,883 and were 
detected in the near-IR \citep{kankare12}. These are likely Type IIP events based on their near-IR 
light-curves and radio upper limits \citep[][\citetalias{rocc12a}]{kankare12}. Type IIP SNe can
easily go unnoticed at radio wavelengths as they typically have low peak luminosities 
($\lesssim 10^{26}$\,\lunits{}). This would imply that the transients we detect \citepalias[reported 
in][and in this paper]{rocc12a}, if they are SNe, are either Type Ib/c (which are bright and evolve fast), 
or are relatively bright Type II's caught close to their maximum light, falling below our detection limits 
rather soon. In both cases, we do not expect to detect them in consecutive yearly epochs.

We do not detect non-thermal milliarcsec radio sources off the IC\,883 nuclear region. However,
we note that there are large amounts of molecular gas throughout the system. The largest concentration 
of gas is located in component B2a \citep[labelled as C1 by][]{zauderer16}, which is also near the 
explosion site of SNe 2010cu and 2011hi \citepalias[see figure 1 in][]{rocc12a}. The molecular mass in 
B2a is about twice as much as in the central component A (i.e., C3). \citet{clemens04} noticed that the 
spectral index in the nuclear region (A) is quite flat (consistent with an AGN nature) and steeper in 
the rest of the system (in agreement with active star formation). It is thus likely that the peak of 
star formation in IC\,883 is off-nuclear \citep[as seen in e.g., the Antennae galaxies,][]{wang04}, 
but yielding a dimmer population of SNe like SNe 2010cu and 2011hi that are not easily
detected at the radio.

\subsection{IC\,883 energetics}\label{sec:energetics} 

The starburst in IC\,883 seems to dominate the global emission at a wide range of wavelengths. 
However, the detection of the mas-scale jet in IC\,883, together with the high brightness 
temperature ($\tb{} > 10^7$\,K) radio core, the measured spectral index and the flux density 
variability, indicate that the central component is non-thermal and related to the presence of an 
AGN, which clearly dominates the nuclear radio emission. This has only been possible owing 
to high-resolution radio observations that allowed us to disentangle the nuclear region. 

There is, however, no observational measurement of the mass of the central BH (\mbh{}) 
in IC\,883. This LIRG is not a classical bulge galaxy nor an elliptical, however, 
\citet{scoville00} found that its 1.6\,\mum{} light profile is better fitted by a $r^{1/4}$ 
law rather than by an exponential disk, thus implying an elliptical-like bulge. It should also
be considered that IC\,883 is an advanced merger, and although $\sigma_{\star}$ can have oscillations 
owing to departures of the system from dynamical relaxation, it is considered that after coalescence, 
as in this case, such fluctuations are not large \citep{stickley14}. We can thus use as a 
good approximation the \mbh{}-$\sigma_{\star}$ relation described in \citet{kormendy13} for 
classical bulges and elliptical galaxies in the IC\,883 case. \citet{rothberg06} measured 
$172\pm8$\,\kms{} in IC\,883 for the central stellar velocity 
dispersion, $\sigma_{\star}$, based on the absorption of the Ca triplet line at 8500\,\AA{}. 
Using this $\sigma_{\star}$ value in equation 7 from \citet{kormendy13}, we estimate 
$\mbh{} \sim (1.6 \pm 0.4)\times 10^{8}$\,\msun{}. In this case, the Eddington Luminosity 
would be within the range  $\ledd \sim (1.5$ -- $2.5)\times 10^{46}$\,\rlunits{}.  

IC\,883 was originally classified as a LINER \citep{veilleux95}. LLAGNs usually 
also have a low-ionisation power, and thus lack coronal lines. The typical Eddington 
factors ($\lemi/\ledd$) of LLAGNs are of the order of $10^{-2}$--$10^{-7}$, which
makes them low-accretion rate AGNs \citep{nagar05}. 

Is IC\,883 a true LLAGN even though a coronal line ([Ne {\small V}] 14.32\,\mum{}) has been
detected in it? To investigate this, we first determine the mass accretion rate for IC\,883. 
For doing this we should calculate $\lemi = \qjet + \lbolagn$, where \qjet{} is the jet 
kinematic luminosity, and \lbolagn{} is the AGN bolometric luminosity. According to equation 
6 from \citet{merloni07}, \qjet{} can be derived from its correlation with the core radio 
luminosity at 5\,GHz. Using our VLBI measurements (Table \ref{tab:lcx_matched}), we estimate 
$\qjet \sim$ (4.4 -- 5.0) $\times 10^{43}$\,\rlunits{}. We also know that the [Ne {\small V}] 
14.32\,\mum{} emission line is detected in IC\,883 with a flux density of 
$(1.61 \pm 0.13) \times 10^{-17}$ W/m$^{2}$ \citep[][]{inami13}, i.e., with a luminosity of 
$\sim (1.8$ -- $2.1) \times 10^{40}$\,\rlunits{} not corrected for extinction. Since the [Ne 
{\small V}] line has an ionisation potential that is too high to be created solely by star 
formation, we can estimate the bolometric AGN luminosity (\lbolagn{}) for the source using 
equation 1 from \citet{satyapal07}, such that $\lbolagn \sim (1.2$ -- $1.4) \times 10^{44}$\,\rlunits{}.
We find that the Eddington factor for the AGN in IC\,883 is $(6.9$ -- $8.5) \times10^{-3}$, and thus,
within the range of values found for low-accretion rate AGNs.

\citet{nagar05} found that the primary accretion energy output for LLAGNs with a compact radio core is
\qjet{}. However, in the case of IC\,883 we find that \lbolagn{} is the main contributor to the emitted 
luminosity. This could indicate that either the mass of the black hole based on the stellar mass 
of the surrounding bulge has been underestimated and/or the [Ne {\small V}] luminosity has an important 
contribution from a young starburst with a significant population of massive stars (Wolf-Rayet or O-type 
stars), as might be the case in optically classified starbursts \citep{abel08}. We note, however, 
that the values we find for \mbh{} and \lbolagn{} fit better within the distribution of normal 
AGNs rather than within the LLAGN category, whereas when looking at the distribution in terms of the
Eddington factor vs. \lbolagn{}, IC\,883 falls at the limit between LLAGNs and normal AGNs 
\citep[figure 1 in][]{elitzur09}. This apparent discrepancy might be related to the fact that IC\,883 
is an advanced merger in its transition to become an AGN dominated source.

Radio observations provide solid evidence for the presence of an AGN, and it is thus reasonable to look 
for an X-ray counterpart, although as in the case of other LIRGs, the AGN in IC\,883 might be highly 
obscured \citep[e.g.,][]{teng15} thus hindering the detection of an AGN contribution at energies lower 
than 10\,keV. Using the \mbh{} value we have estimated from the bulge-mass relationship, we can infer 
the X-ray luminosity of the AGN by means of the fundamental plane of BH activity \citep[][equation 
5]{merloni03}. We obtain $L_{2-10\rmn{keV}} \simeq (2.7$ -- $3.6) \times 10^{42}$\,\rlunits{}, which is 
significantly larger than the total observed luminosity. One plausible explanation is that the AGN 
is obscured by Compton-thick ($\geq 10^{24}$\,cm$^{-2}$) material, as has been suggested previously 
by \citet{zauderer16} based on the observed  high column densities, and by \citet{asmus15} based
on the Mid-IR--X-ray correlation. Assuming a column density of $10^{25}$\,cm$^{-2}$, the AGN observed 
luminosity would decrease down to $10^{40}$\,\rlunits{}. This corresponds to a very obscured AGN contributing 
$\sim$10 per cent of the total observed X-ray luminosity we reported in \S \ref{sec:xray}. Compton-thick 
AGNs are rather common in the local Universe, and are believed to represent $\sim$20--30 per cent of the 
whole AGN population \citep{burlon11,ricci15}. Hence, observations at energies $>10$\,keV can help to unveil 
a heavily obscured AGN in the system, as done for NGC\,6286 \citep{ricci16}.

It is possible now to calculate the IR luminosity of the dust around the AGN from the unobscured 2--10\,keV 
luminosity. For this, we use the empirical relationship for the AGN thermal emission presented by 
\citet{mullaney11}. We obtain a luminosity of $\liragn \simeq (1.2$ -- 2.5$)\times 10^{43}$\rlunits{}. 
Comparing this value with the IR luminosity of the system ($\lir \simeq 1.8 \times 10^{45}$\rlunits{}), 
we find that the AGN contribution to the total IR luminosity is below two per cent, consistent with the 
SED fitting presented in \citetalias{rocc12a}.

\subsection{The starburst-AGN connection}

The properties that the AGN in IC\,883 has in common with GPS sources are indicative of 
a short age ($\lesssim 10^4$\,yr). If the AGN in IC\,883 was more powerful in the past and 
produced kpc-scale jets, we should be able to detect them in the MHz regime with  e.g., the 
LOw Fequency ARray (LOFAR). \citet{shulevski15} reported the case of a radio galaxy within 
the Abell\,407 cluster, which is characterised by a steep spectrum all the way from $<100$\,MHz 
to a few GHz in frequency and displays kpc-scale radio lobes, and thus shows evidence for past 
AGN activity. Indeed, GPS sources have rarely been found in connection with relic emission 
\citep{orienti15}. The observed GPS-like spectrum in IC\,883, highly inverted at low frequencies, 
and the absence of large scale jet emission, indicate that this scenario is not plausible and the 
GPS-like core should be young. This would also explain why we do not see more evidence of it at 
other wavelengths. For instance, \citet{stierwalt13} argued that there is a delay of a few hundred 
million years for the AGN to dominate the mid-IR emission after the merger has taken place.  

The AGN is apparently in an early stage in its evolution, in agreement with a GPS scenario. From 
the infrared SED fitting in \citetalias{rocc12a}, the age of the starburst would be 55\,Myr.
Although estimates of starburst age from radiative transfer model fits to the infrared SED are 
uncertain, we note that the Mid-IR spectrum displays very strong PAH features which in the 
\citet{efstathiou00} model is an indication of a fairly old age. This agrees well with the 
lack of a dense molcular gas component reported by \citet{papadopoulos14}. The starburst is 
thus much older than the putative age of the AGN core, and therefore, it is reasonable to 
assume at least that the AGN did not trigger the starburst. 

\section{Conclusions}\label{sec:fin} 

We performed three yearly epochs of multi-frequency EVN observations toward IC\,883. Our $<1$\,mas 
resolution observations at 8.4\,GHz reveal for the first time a core-jet morphology in the nucleus 
of IC\,883. The parsec-scale jet is subluminal (0.6--1\,$c$), similar to those found in other LLAGNs.
However, IC\,883 displays more properties in common with normal AGNs.

X-ray analysis is inconclusive about the origin of this emission, either AGN or SF.
To estimate the contribution of an AGN to the X-ray luminosity, we resort to the use of the 
black hole mass obtained by means of its correlation with the mass of the bulge. We estimated
$\mbh{} \sim (1.6 \pm 0.4)\times 10^{8}$\,\msun{}. This is only slightly less massive than 
the radio-loud AGN in the host of the second closest GPS source \citep[IC\,1459;][]{tingay15}. 
Plugging in the \mbh{} value in the fundamental plane of BH activity, we find that the AGN in 
IC\,883 would have a $L_{2-10\rmn{keV}} \simeq (2.7$ -- $3.6) \times 10^{42}$\,\rlunits{}, with a contribution 
to the observed luminosity down to ten per cent if it is a Compton-thick AGN, as is likely the case. 
Observations at higher energies are needed to corroborate our estimates. 

The star formation is clearly powering the energetics of IC\,883 at all wavelengths, with the 
AGN contributing only less than two per cent of the total IR luminosity. At radio wavelengths however,
an AGN is dominating the nuclear emission. The apparent discrepancies among the interpretation 
of different diagnostic tools could simply be pointing to the transitional nature of IC\,883 from a 
very active star formation phase, to an AGN dominated one which at the moment is presumably in an 
early phase.  

The radio SED of the IC\,883 core is characterised by a highly inverted spectral index ($\sim2.5$) with 
a turnover at low frequencies ($\sim4.4$\,GHz). This fact, together with the compactness of the nucleus,
make IC\,883 a candidate for a GPS source embedded in an 
advanced merger which remarkably is still actively forming stars. Based on its radio luminosity and
linear size, we determine a magnetic field strength of $\sim2.6$\,mG, which agrees with typical values for
GPS sources, but being significantly larger than  for advanced mergers. Correspondingly,
we estimate an approximate age for the source of 3$\times10^3$\,yr. Together with the galaxies studied
by \citet{tingay03}, IC\,883 is potentially one of the nearest and youngest GPS galaxies and by far 
the least luminous one, being three orders of magnitude less luminous than is typical. 

From the available observations, we infer that the radio SED of the IC\,883 core is quite flat over a 
wide frequency range (8.4 -- 32.5\,GHz). IC\,883 could then be classified as a mere flat-spectrum source 
rather than a GPS \citep{snellen99}. However, the flat-spectrum sources to which \citeauthor{snellen99} 
refer are quasars and BL LACs, which are known to be highly variable. In this case, although there is
some flux density variability, the spectrum is quite similar in both 2013 and 2014. We note that 
\citet{torniainen05} sampled the SED of several bona-fide GPS sources and candidates and found at least 
two within the GPS category which show flat spectral indices over a wide frequency range, similar to what 
IC\,883 is displaying. The sampling of the SED at higher frequencies is certainly missing and currently there 
is no available information on the optically-thin part of the spectrum. Once this is attained, we will be 
able to confirm or rule out the GPS nature of the IC\,883 core. 

Optical imaging has proved that there must be a relation between merging/interacting systems and 
GPS sources \citep{stanghellini93}. However, it is rather uncommon that GPS sources are hosted by active 
star-forming galaxies. In fact, there is to our knowledge only one more case in the literature. 
\citet{norris12} reported on a radio-loud  
AGN with kpc-scale jets and radio spectra similar to that of GPS sources, embedded in a star-forming 
ULIRG. IC\,883 is also hosted by a LIRG galaxy, but remains the least luminous GPS source, and the 
youngest one ever found in a very active star forming environment. GPS sources have been more
easily identified at high redshifts and current samples are biased toward high luminosity sources 
\citep{odea98}. The low-luminosity population remains relatively unknown and poorly sampled.

Our study thus opens a new window to investigate the physical environments of low-luminosity, 
young GPS sources, which can potentially be a numerous population among LIRGs in an advanced 
merger stage, but have so far been missed presumably owing to the lack of high-resolution, high-sensitivity
radio data. 

%%%%%%%%%%%%%%%%%%%%%%%%%%%%%%%%%%%%%%%%%%%%%%%%%% ACKNOWLEDGEMENTS
 \footnotesize{
\section*{Acknowledgements}
The authors thank George Privon and Jos\'e Luis Prieto for constructive comments on the manuscript,
Xian Chen and Mar Mezcua for insightful discussions, and the anonymous referee for providing detailed 
comments that have improved our manuscript. We acknowledge financial support by the Ministry of Economy, 
Development, and Tourism's Millennium Science Initiative through grant IC120009, awarded to The Millennium 
Institute of Astrophysics, MAS, Chile (CR-C, FEB), from CONICYT through FONDECYT grants 3150238 (CR-C), 
1140304 (PA), 1141218 and 1151408, from the CONICYT-Chile grants ``EMBIGGEN" Anillo ACT1101, Basal-CATA 
PFB--06/2007 and the China-CONICYT fellowship (CR, FEB), as well as funding from the Spanish Ministry 
of Economy and Competitiveness under grants AYA2012-38491-CO2-02 and AYA2015-63939-C2-1-P, which are partly 
funded by the FEDER programme (AA, MAP-T, RH-I). The European VLBI Network is a joint facility of 
independent European, African, Asian, and North American radio astronomy institutes. Scientific results from 
data presented in this publication are derived from the EVN project code ER030. The research leading to these 
results has received funding from the European Commission Seventh Framework Programme (FP/2007-2013) 
under grant agreement No. 283393 (RadioNet3). The National Radio Astronomy Observatory is a 
facility of the National Science Foundation operated under cooperative agreement by Associated 
Universities, Inc.
}
%%%%%%%%%%%%%%%%%%%%%%%%%%%%%%%%%%%%%%%%%%%%%%%%%% BIBLIOGRAPHY
 \footnotesize{

}
%%%%%%%%%%%%%%%%%%%%%%%%%%%%%%%%%%%%%%%%%%%%%%%%%% END OF PAPER

% Don't change these lines
%\bsp	% typesetting comment
\label{lastpage}


\begin{thebibliography}{89}
\expandafter\ifx\csname natexlab\endcsname\relax\def\natexlab#1{#1}\fi

\bibitem[{Abel} \& {Satyapal}(2008)]{abel08}
{Abel} N.~P., {Satyapal} S., 2008, \apj, 678, 686

\bibitem[{Argo} et~al.(2015){Argo}, {van Bemmel}, {Connolly} \&
  {Beswick}]{argo15}
{Argo} M.~K., {van Bemmel} I.~M., {Connolly} S.~D., {Beswick} R.~J., 2015,
  \mnras, 452, 1081

\bibitem[{Armus} et~al.(2009){Armus}, {Mazzarella}, {Evans} et~al.]{armus09}
{Armus} L., {Mazzarella} J.~M., {Evans} A.~S., et~al., 2009, \pasp, 121, 559

\bibitem[{Asmus} et~al.(2015){Asmus}, {Gandhi}, {H{\"o}nig}, {Smette} \&
  {Duschl}]{asmus15}
{Asmus} D., {Gandhi} P., {H{\"o}nig} S.~F., {Smette} A., {Duschl} W.~J., 2015,
  \mnras, 454, 766

\bibitem[{Barcos-Mu{\~n}oz} et~al.(2016){Barcos-Mu{\~n}oz}, {Leroy}, {Evans} \&
  {et al.}]{barcos16}
{Barcos-Mu{\~n}oz} L., {Leroy} A., {Evans} A., {et al.}, 2016, in { The
  Interplay between Local and Global Processes in Galaxies,\/}

\bibitem[{Beck} \& {Krause}(2005)]{beck05}
{Beck} R., {Krause} M., 2005, Astronomische Nachrichten, 326, 414

\bibitem[{Bicknell} et~al.(1997){Bicknell}, {Dopita} \& {O'Dea}]{bicknell97}
{Bicknell} G.~V., {Dopita} M.~A., {O'Dea} C.~P.~O., 1997, \apj, 485, 112

\bibitem[{Bicknell} et~al.(2003){Bicknell}, {Saxton}, {Sutherland}, {Midgley}
  \& {Wagner}]{bicknell03}
{Bicknell} G.~V., {Saxton} C.~J., {Sutherland} R.~S., {Midgley} S., {Wagner}
  S.~J., 2003, \nar, 47, 537

\bibitem[{Burlon} et~al.(2011){Burlon}, {Ajello}, {Greiner}, {Comastri},
  {Merloni} \& {Gehrels}]{burlon11}
{Burlon} D., {Ajello} M., {Greiner} J., {Comastri} A., {Merloni} A., {Gehrels}
  N., 2011, \apj, 728, 58

\bibitem[{Callingham} et~al.(2015){Callingham}, {Gaensler}, {Ekers}
  et~al.]{callingham15}
{Callingham} J.~R., {Gaensler} B.~M., {Ekers} R.~D., et~al., 2015, \apj, 809,
  168

\bibitem[{Carrera} et~al.(2007){Carrera}, {Ebrero}, {Mateos} et~al.]{carrera07}
{Carrera} F.~J., {Ebrero} J., {Mateos} S., et~al., 2007, \aap, 469, 27

\bibitem[{Clemens} \& {Alexander}(2004)]{clemens04}
{Clemens} M.~S., {Alexander} P., 2004, \mnras, 350, 66

\bibitem[{Condon}(1992)]{condon92}
{Condon} J.~J., 1992, \araa, 30, 575

\bibitem[{Condon} et~al.(1991){Condon}, {Huang}, {Yin} \& {Thuan}]{condon91}
{Condon} J.~J., {Huang} Z.-P., {Yin} Q.~F., {Thuan} T.~X., 1991, \apj, 378, 65

\bibitem[{Conway}(1999)]{conway99}
{Conway} J.~E., 1999, \nar, 43, 509

\bibitem[{de Vries} et~al.(1997){de Vries}, {Barthel} \& {O'Dea}]{devries97}
{de Vries} W.~H., {Barthel} P.~D., {O'Dea} C.~P., 1997, \aap, 321, 105

\bibitem[{Di Matteo} et~al.(2007){Di Matteo}, {Combes}, {Melchior} \&
  {Semelin}]{dimatteo07}
{Di Matteo} P., {Combes} F., {Melchior} A.-L., {Semelin} B., 2007, \aap, 468,
  61

\bibitem[{Downes} \& {Solomon}(1998)]{downes98}
{Downes} D., {Solomon} P.~M., 1998, \apj, 507, 615

\bibitem[{Drzazga} et~al.(2011){Drzazga}, {Chy{\.z}y}, {Jurusik} \&
  {Wi{\'o}rkiewicz}]{drzazga11}
{Drzazga} R.~T., {Chy{\.z}y} K.~T., {Jurusik} W., {Wi{\'o}rkiewicz} K., 2011,
  \aap, 533, A22

\bibitem[{Dudik} et~al.(2009){Dudik}, {Satyapal} \& {Marcu}]{dudik09}
{Dudik} R.~P., {Satyapal} S., {Marcu} D., 2009, \apj, 691, 1501

\bibitem[{Efstathiou} et~al.(2000){Efstathiou}, {Rowan-Robinson} \&
  {Siebenmorgen}]{efstathiou00}
{Efstathiou} A., {Rowan-Robinson} M., {Siebenmorgen} R., 2000, \mnras, 313, 734

\bibitem[{Elitzur} \& {Ho}(2009)]{elitzur09}
{Elitzur} M., {Ho} L.~C., 2009, \apjl, 701, L91

\bibitem[{Fanali} et~al.(2013){Fanali}, {Caccianiga}, {Severgnini}
  et~al.]{fanali13}
{Fanali} R., {Caccianiga} A., {Severgnini} P., et~al., 2013, \mnras, 433, 648

\bibitem[{Garmire} et~al.(2003){Garmire}, {Bautz}, {Ford}, {Nousek} \&
  {Ricker}]{garmire03}
{Garmire} G.~P., {Bautz} M.~W., {Ford} P.~G., {Nousek} J.~A., {Ricker} Jr.
  G.~R., 2003, in { X-Ray and Gamma-Ray Telescopes and Instruments for
  Astronomy.\/}, edited by J.~E. {Truemper}, H.~D. {Tananbaum}, vol. 4851 of {
  \procspie\/},  28--44

\bibitem[{Hancock} et~al.(2009){Hancock}, {Tingay}, {Sadler}, {Phillips} \&
  {Deller}]{hancock09}
{Hancock} P.~J., {Tingay} S.~J., {Sadler} E.~M., {Phillips} C., {Deller} A.~T.,
  2009, \mnras, 397, 2030

\bibitem[{Hopkins} et~al.(2006){Hopkins}, {Hernquist}, {Cox}, {Di Matteo},
  {Robertson} \& {Springel}]{hopkins06}
{Hopkins} P.~F., {Hernquist} L., {Cox} T.~J., {Di Matteo} T., {Robertson} B.,
  {Springel} V., 2006, \apjs, 163, 1

\bibitem[{Inami} et~al.(2013){Inami}, {Armus}, {Charmandaris} et~al.]{inami13}
{Inami} H., {Armus} L., {Charmandaris} V., et~al., 2013, \apj, 777, 156

\bibitem[{Iwasawa} et~al.(2011){Iwasawa}, {Sanders}, {Teng} et~al.]{iwasawa11}
{Iwasawa} K., {Sanders} D.~B., {Teng} S.~H., et~al., 2011, \aap, 529, A106

\bibitem[{Jansen} et~al.(2001){Jansen}, {Lumb}, {Altieri} et~al.]{jansen01}
{Jansen} F., {Lumb} D., {Altieri} B., et~al., 2001, \aap, 365, L1

\bibitem[{Kankare} et~al.(2012){Kankare}, {Mattila}, {Ryder} et~al.]{kankare12}
{Kankare} E., {Mattila} S., {Ryder} S., et~al., 2012, \apjl, 744, L19

\bibitem[{Kormendy} \& {Ho}(2013)]{kormendy13}
{Kormendy} J., {Ho} L.~C., 2013, \araa, 51, 511

\bibitem[{Lonsdale} et~al.(1993){Lonsdale}, {Smith} \& {Lonsdale}]{lonsdale93}
{Lonsdale} C.~J., {Smith} H.~J., {Lonsdale} C.~J., 1993, \apjl, 405, L9

\bibitem[{McMullin} et~al.(2007){McMullin}, {Waters}, {Schiebel}, {Young} \&
  {Golap}]{mcmullin07}
{McMullin} J.~P., {Waters} B., {Schiebel} D., {Young} W., {Golap} K., 2007, in
  { Astronomical Data Analysis Software and Systems XVI\/}, edited by R.~A.
  {Shaw}, F.~{Hill}, D.~J. {Bell}, vol. 376 of { Astronomical Society of the
  Pacific Conference Series\/},  127

\bibitem[{Merloni} \& {Heinz}(2007)]{merloni07}
{Merloni} A., {Heinz} S., 2007, \mnras, 381, 589

\bibitem[{Merloni} et~al.(2003){Merloni}, {Heinz} \& {di Matteo}]{merloni03}
{Merloni} A., {Heinz} S., {di Matteo} T., 2003, \mnras, 345, 1057

\bibitem[{Mezcua} \& {Prieto}(2014)]{mezcua14}
{Mezcua} M., {Prieto} M.~A., 2014, \apj, 787, 62

\bibitem[{Modica} et~al.(2012){Modica}, {Vavilkin}, {Evans} et~al.]{modica12}
{Modica} F., {Vavilkin} T., {Evans} A.~S., et~al., 2012, \aj, 143, 16

\bibitem[{Momjian} et~al.(2003){Momjian}, {Romney}, {Carilli}, {Troland} \&
  {Taylor}]{momjian03}
{Momjian} E., {Romney} J.~D., {Carilli} C.~L., {Troland} T.~H., {Taylor} G.~B.,
  2003, \apj, 587, 160

\bibitem[{Mullaney} et~al.(2011){Mullaney}, {Alexander}, {Goulding} \&
  {Hickox}]{mullaney11}
{Mullaney} J.~R., {Alexander} D.~M., {Goulding} A.~D., {Hickox} R.~C., 2011,
  \mnras, 414, 1082

\bibitem[{Murgia}(2003)]{murgia03}
{Murgia} M., 2003, \pasa, 20, 19

\bibitem[{Nagar} et~al.(2005){Nagar}, {Falcke} \& {Wilson}]{nagar05}
{Nagar} N.~M., {Falcke} H., {Wilson} A.~S., 2005, \aap, 435, 521

\bibitem[{Nagar} et~al.(2002{\natexlab{a}}){Nagar}, {Falcke}, {Wilson} \&
  {Ulvestad}]{nagar02b}
{Nagar} N.~M., {Falcke} H., {Wilson} A.~S., {Ulvestad} J.~S.,
  2002{\natexlab{a}}, \aap, 392, 53

\bibitem[{Nagar} et~al.(2002{\natexlab{b}}){Nagar}, {Wilson}, {Falcke},
  {Ulvestad} \& {Mundell}]{nagar02a}
{Nagar} N.~M., {Wilson} A.~S., {Falcke} H., {Ulvestad} J.~S., {Mundell} C.~G.,
  2002{\natexlab{b}}, in { Issues in Unification of Active Galactic Nuclei\/},
  edited by R.~{Maiolino}, A.~{Marconi}, N.~{Nagar}, vol. 258 of { Astronomical
  Society of the Pacific Conference Series\/},  171

\bibitem[{Norris} et~al.(2012){Norris}, {Lenc}, {Roy} \& {Spoon}]{norris12}
{Norris} R.~P., {Lenc} E., {Roy} A.~L., {Spoon} H., 2012, \mnras, 422, 1453

\bibitem[{O'Dea}(1998)]{odea98}
{O'Dea} C.~P., 1998, \pasp, 110, 493

\bibitem[{O'Dea} et~al.(1991){O'Dea}, {Baum} \& {Stanghellini}]{odea91}
{O'Dea} C.~P., {Baum} S.~A., {Stanghellini} C., 1991, \apj, 380, 66

\bibitem[{Orienti}(2015)]{orienti15}
{Orienti} M., 2015, ArXiv e-prints

\bibitem[{Pacholczyk}(1970)]{pacholczyk}
{Pacholczyk} A.~G., 1970, {Radio Astrophysics: Nonthermal processes in galactic
  and extragalactic sources}, W. H. Freeman and Company

\bibitem[{Papadopoulos} et~al.(2014){Papadopoulos}, {Zhang}, {Xilouris}
  et~al.]{papadopoulos14}
{Papadopoulos} P.~P., {Zhang} Z.-Y., {Xilouris} E.~M., et~al., 2014, \apj, 788,
  153

\bibitem[{Parra} et~al.(2010){Parra}, {Conway}, {Aalto} et~al.]{parra10}
{Parra} R., {Conway} J.~E., {Aalto} S., et~al., 2010, \apj, 720, 555

\bibitem[{Polatidis} \& {Conway}(2003)]{polatidis03}
{Polatidis} A.~G., {Conway} J.~E., 2003, \pasa, 20, 69

\bibitem[{Ranalli} et~al.(2003){Ranalli}, {Comastri} \& {Setti}]{ranalli03}
{Ranalli} P., {Comastri} A., {Setti} G., 2003, \aap, 399, 39

\bibitem[{Ricci} et~al.(2016){Ricci}, {Bauer}, {Treister} et~al.]{ricci16}
{Ricci} C., {Bauer} F.~E., {Treister} E., et~al., 2016, \apj, 819, 4

\bibitem[{Ricci} et~al.(2015){Ricci}, {Ueda}, {Koss}, {Trakhtenbrot}, {Bauer}
  \& {Gandhi}]{ricci15}
{Ricci} C., {Ueda} Y., {Koss} M.~J., {Trakhtenbrot} B., {Bauer} F.~E., {Gandhi}
  P., 2015, \apjl, 815, L13

\bibitem[{Rich} et~al.(2014){Rich}, {Kewley} \& {Dopita}]{rich14}
{Rich} J.~A., {Kewley} L.~J., {Dopita} M.~A., 2014, \apjl, 781, L12

\bibitem[{Romero-Ca{\~n}izales}
  et~al.(2012{\natexlab{a}}){Romero-Ca{\~n}izales}, {P{\'e}rez-Torres} \&
  {Alberdi}]{rocc12b}
{Romero-Ca{\~n}izales} C., {P{\'e}rez-Torres} M.~{\'A}., {Alberdi} A.,
  2012{\natexlab{a}}, \mnras, 422, 510

\bibitem[{Romero-Ca{\~n}izales}
  et~al.(2012{\natexlab{b}}){Romero-Ca{\~n}izales}, {P{\'e}rez-Torres},
  {Alberdi} et~al.]{rocc12a}
{Romero-Ca{\~n}izales} C., {P{\'e}rez-Torres} M.~A., {Alberdi} A., et~al.,
  2012{\natexlab{b}}, \aap, 543, A72

\bibitem[{Rothberg} \& {Joseph}(2006)]{rothberg06}
{Rothberg} B., {Joseph} R.~D., 2006, \aj, 131, 185

\bibitem[{Sanders} et~al.(2003){Sanders}, {Mazzarella}, {Kim}, {Surace} \&
  {Soifer}]{sanders03}
{Sanders} D.~B., {Mazzarella} J.~M., {Kim} D.-C., {Surace} J.~A., {Soifer}
  B.~T., 2003, \aj, 126, 1607

\bibitem[{Sanders} \& {Mirabel}(1996)]{sanders96}
{Sanders} D.~B., {Mirabel} I.~F., 1996, \araa, 34, 749

\bibitem[{Satyapal} et~al.(2007){Satyapal}, {Vega}, {Heckman}, {O'Halloran} \&
  {Dudik}]{satyapal07}
{Satyapal} S., {Vega} D., {Heckman} T., {O'Halloran} B., {Dudik} R., 2007,
  \apjl, 663, L9

\bibitem[{Scoville} et~al.(2000){Scoville}, {Evans}, {Thompson}
  et~al.]{scoville00}
{Scoville} N.~Z., {Evans} A.~S., {Thompson} R., et~al., 2000, \aj, 119, 991

\bibitem[{Shepherd} et~al.(1995){Shepherd}, {Pearson} \& {Taylor}]{shepherd95}
{Shepherd} M.~C., {Pearson} T.~J., {Taylor} G.~B., 1995, in { Bulletin of the
  American Astronomical Society\/}, edited by {B.~J.~Butler \& D.~O.~Muhleman},
  vol.~27 of { Bulletin of the American Astronomical Society\/},  903--+

\bibitem[{Shulevski} et~al.(2015){Shulevski}, {Morganti}, {Barthel}
  et~al.]{shulevski15}
{Shulevski} A., {Morganti} R., {Barthel} P.~D., et~al., 2015, \aap, 579, A27

\bibitem[{Smith} et~al.(1995){Smith}, {Herter}, {Haynes}, {Beichman} \&
  {Gautier}]{smith95}
{Smith} D.~A., {Herter} T., {Haynes} M.~P., {Beichman} C.~A., {Gautier} III
  T.~N., 1995, \apj, 439, 623

\bibitem[{Smith} et~al.(1998){Smith}, {Lonsdale} \& {Lonsdale}]{smith98}
{Smith} H.~E., {Lonsdale} C.~J., {Lonsdale} C.~J., 1998, \apj, 492, 137

\bibitem[{Smith} et~al.(2001){Smith}, {Brickhouse}, {Liedahl} \&
  {Raymond}]{smith01}
{Smith} R.~K., {Brickhouse} N.~S., {Liedahl} D.~A., {Raymond} J.~C., 2001,
  \apjl, 556, L91

\bibitem[{Snellen}(2009)]{snellen09}
{Snellen} I., 2009, in { Approaching Micro-Arcsecond Resolution with VSOP-2:
  Astrophysics and Technologies\/}, edited by Y.~{Hagiwara}, E.~{Fomalont},
  M.~{Tsuboi}, M.~{Yasuhiro}, vol. 402 of { Astronomical Society of the Pacific
  Conference Series\/},  221

\bibitem[{Snellen} et~al.(1999){Snellen}, {Schilizzi}, {Miley}, {Bremer},
  {R{\"o}ttgering} \& {van Langevelde}]{snellen99}
{Snellen} I.~A.~G., {Schilizzi} R.~T., {Miley} G.~K., {Bremer} M.~N.,
  {R{\"o}ttgering} H.~J.~A., {van Langevelde} H.~J., 1999, \nar, 43, 675

\bibitem[{Snellen} et~al.(2000){Snellen}, {Schilizzi}, {Miley}, {de Bruyn},
  {Bremer} \& {R{\"o}ttgering}]{snellen00}
{Snellen} I.~A.~G., {Schilizzi} R.~T., {Miley} G.~K., {de Bruyn} A.~G.,
  {Bremer} M.~N., {R{\"o}ttgering} H.~J.~A., 2000, \mnras, 319, 445

\bibitem[{Spoon} et~al.(2007){Spoon}, {Marshall}, {Houck} et~al.]{spoon07}
{Spoon} H.~W.~W., {Marshall} J.~A., {Houck} J.~R., et~al., 2007, \apjl, 654,
  L49

\bibitem[{Stanghellini} et~al.(1993){Stanghellini}, {O'Dea}, {Baum} \&
  {Laurikainen}]{stanghellini93}
{Stanghellini} C., {O'Dea} C.~P., {Baum} S.~A., {Laurikainen} E., 1993, \apjs,
  88, 1

\bibitem[{Stickley} \& {Canalizo}(2014)]{stickley14}
{Stickley} N.~R., {Canalizo} G., 2014, \apj, 786, 12

\bibitem[{Stierwalt} et~al.(2013){Stierwalt}, {Armus}, {Surace}
  et~al.]{stierwalt13}
{Stierwalt} S., {Armus} L., {Surace} J.~A., et~al., 2013, \apjs, 206, 1

\bibitem[{Str{\"u}der} et~al.(2001){Str{\"u}der}, {Briel}, {Dennerl}
  et~al.]{struder01}
{Str{\"u}der} L., {Briel} U., {Dennerl} K., et~al., 2001, \aap, 365, L18

\bibitem[{Teng} et~al.(2015){Teng}, {Rigby}, {Stern} et~al.]{teng15}
{Teng} S.~H., {Rigby} J.~R., {Stern} D., et~al., 2015, \apj, 814, 56

\bibitem[{Tingay} \& {Edwards}(2015)]{tingay15}
{Tingay} S.~J., {Edwards} P.~G., 2015, \mnras, 448, 252

\bibitem[{Tingay} et~al.(2003){Tingay}, {Edwards} \& {Tzioumis}]{tingay03}
{Tingay} S.~J., {Edwards} P.~G., {Tzioumis} A.~K., 2003, \mnras, 346, 327

\bibitem[{Tingay} et~al.(1997){Tingay}, {Jauncey}, {Reynolds} et~al.]{tingay97}
{Tingay} S.~J., {Jauncey} D.~L., {Reynolds} J.~E., et~al., 1997, \aj, 113, 2025

\bibitem[{Torniainen} et~al.(2005){Torniainen}, {Tornikoski}, {Ter{\"a}sranta},
  {Aller} \& {Aller}]{torniainen05}
{Torniainen} I., {Tornikoski} M., {Ter{\"a}sranta} H., {Aller} M.~F., {Aller}
  H.~D., 2005, \aap, 435, 839

\bibitem[{Turner} et~al.(2001){Turner}, {Abbey}, {Arnaud} et~al.]{turner01}
{Turner} M.~J.~L., {Abbey} A., {Arnaud} M., et~al., 2001, \aap, 365, L27

\bibitem[{Tyul'Bashev}(2001)]{tyulbashev01}
{Tyul'Bashev} S.~A., 2001, Astronomy Reports, 45, 428

\bibitem[{van der Laan} \& {Perola}(1969)]{vanderlaan69}
{van der Laan} H., {Perola} G.~C., 1969, \aap, 3, 468

\bibitem[{Veilleux} et~al.(1995){Veilleux}, {Kim}, {Sanders}, {Mazzarella} \&
  {Soifer}]{veilleux95}
{Veilleux} S., {Kim} D.-C., {Sanders} D.~B., {Mazzarella} J.~M., {Soifer}
  B.~T., 1995, \apjs, 98, 171

\bibitem[{Wang} et~al.(2004){Wang}, {Fazio}, {Ashby} et~al.]{wang04}
{Wang} Z., {Fazio} G.~G., {Ashby} M.~L.~N., et~al., 2004, \apjs, 154, 193

\bibitem[{Weisskopf} et~al.(2000){Weisskopf}, {Tananbaum}, {Van Speybroeck} \&
  {O'Dell}]{weisskopf00}
{Weisskopf} M.~C., {Tananbaum} H.~D., {Van Speybroeck} L.~P., {O'Dell} S.~L.,
  2000, in { X-Ray Optics, Instruments, and Missions III\/}, edited by J.~E.
  {Truemper}, B.~{Aschenbach}, vol. 4012 of { \procspie\/},  2--16

\bibitem[{Wilms} et~al.(2000){Wilms}, {Allen} \& {McCray}]{wilms00}
{Wilms} J., {Allen} A., {McCray} R., 2000, \apj, 542, 914

\bibitem[{Yuan} et~al.(2010){Yuan}, {Kewley} \& {Sanders}]{yuan10}
{Yuan} T.-T., {Kewley} L.~J., {Sanders} D.~B., 2010, \apj, 709, 884

\bibitem[{Zauderer} et~al.(2016){Zauderer}, {Bolatto}, {Vogel}
  et~al.]{zauderer16}
{Zauderer} B.~A., {Bolatto} A.~D., {Vogel} S.~N., et~al., 2016, \aj, 151, 18

\end{thebibliography}
\end{document}